\documentclass[10pt]{iopart}

\usepackage{booktabs}
\usepackage{graphicx}
\usepackage{iopams}
\usepackage{lipsum}
\usepackage{orcidlink}
\usepackage{subcaption}

\definecolor{darkgreen}{RGB}{0,170,0}

\newcommand{\brems}{brems\-strah\-lung}

\newcommand{\siname}{Supplementary Material}

\begin{document}


    \title[Stand-off runaway termination by tungsten particulates]{Stand-off runaway electron beam termination by tungsten particulates for tokamak disruption mitigation}

    \author{Michael A. Lively,$^1$  \orcidlink{0000-0001-6511-9852}
        Danny Perez,$^1$        \orcidlink{0000-0003-3028-5249}
        Blas P. Uberuaga,$^2$   \orcidlink{0000-0001-6934-6219}
        Yanzeng Zhang$^1$       \orcidlink{0000-0002-1856-2701}
        and Xian-Zhu Tang$^1$   \orcidlink{0000-0002-4036-6643}}

    \address{$^1$ Theoretical Division, Los Alamos National Laboratory, Los Alamos, NM 87545, United States of America}
    \address{$^2$ Materials Science and Technology Division, Los Alamos National Laboratory, Los Alamos, NM 87545, United States of America}
    \ead{livelym@lanl.gov}
    \ead{xtang@lanl.gov}

    \begin{abstract}
        Stand-off runaway electron termination by injected tungsten particulates offers a plausible option in the toolbox of disruption mitigation. Tungsten is an attractive material choice for this application due to large electron stopping power and high melting point. To assess the feasibility of this scheme, we simulate runaway collisions with tungsten particulates using the MCNP program for incident runaway energies ranging from 1 to 10 MeV. We assess runaway termination from energetics and collisional kinematics perspectives. Energetically, the simulations show that 99\% of runaway beam energy is removed by tungsten particulates on a timescale of 4--9 $\mu$s. Kinematically, the simulations show that 99\% of runaways are terminated by absorption or backscattering on a timescale of 3--4 $\mu$s. By either metric, the runaway beam is effectively terminated before the onset of particulate melting. Furthermore, the simulations show that secondary radiation emission by tungsten particulates does not significantly impact the runaway termination efficacy of this scheme. Secondary radiation is emitted at lower particle energies than the incident runaways and with a broad angular distribution such that the majority of secondary electrons emitted will not experience efficient runaway re-acceleration. Overall, the stand-off runaway termination scheme is a promising concept for last-ditch runaway mitigation in ITER, SPARC, and other future burning-plasma tokamaks.
    \end{abstract}

    \noindent \textit{Keywords\/:} runaway electrons, tungsten particulates, disruption mitigation, Monte Carlo

    \submitto{\NF}

    \ioptwocol


    \section{Introduction}
    \label{sec:intro}

    In high-plasma-current tokamak discharges, such as ITER \cite{ITER-2018} and SPARC \cite{RodriguezFernandez-2022-NuclFus}, major disruptions can lead to Ohmic-to-runaway current conversion \cite{Martin-Solis-2017-NuclFus,Breizman-2019-NuclFus,McDevitt-2019-PPCF,Vallhagen-2020-JPlasPhys} through the highly efficient avalanche mechanism \cite{Rosenbluth-1997-NuclFus}. Runaway impact on the vessel wall can melt the first wall and damage or destroy underlying reactor components~\cite{Reux-2015-NuclFus,Matthews-2016-PhysScr,Coburn-2022-NuclFus}. Therefore, safely terminating runaway current is a critical challenge in tokamak disruption mitigation.

    During tokamak operations under nominal conditions, the bulk thermal electrons ($T_\mathrm{e} \approx 15~\mathrm{keV}$) carry the plasma current. However, under disruption conditions \cite{Lehnen-2015-JNuclMater} a minority population of supra-thermal electrons may carry most of the plasma current. This minority population experiences runaway acceleration along magnetic field lines by the parallel electric field to $\gtrsim$MeV energies, thus becoming runaway electrons. Initially, these runaways orbit the tokamak aligned with the magnetic field lines. However, as the magnetic field geometry shifts during a disruption event, such as a vertical displacement event (VDE), the runaway beam contacts the wall and deposits its energy into a surface area of 10 to 100 cm$^2$ \cite{Matthews-2016-PhysScr}, melting the first wall and possibly damaging, e.g., the subsurface cooling system components.

    Current approaches for runaway termination include: (1) massive high-$Z$ impurity injection to enhance runaway current dissipation \cite{Reux-2015-NuclFus,Shiraki-2018-NuclFus}, (2) runaway flush by self-excited major magneto\-hydro\-dynamic (MHD) instabilities after purging of high-$Z$ impurities \cite{Reux-2021-PhysRevLett,PazSoldan-2021-NuclFus,McDevitt-2023-PhysRevE}, and (3) enhanced runaway loss and current dissipation via field line stochasticization by inductively-driven helical field coils \cite{Boozer-2011-PPCF,Izzo-2022-NuclFus}. In each case, the goal is to spread the runaway final loss load on the wall as broadly as possible. In the event that adequate broadening of the runaway load is not possible, a last-ditch defense against significant first wall damage is necessary, such as installing additional armor or sacrificial limiters on the first wall where runaway impact can be reliably predicted \cite{Maviglia-2022-FusEngDes}.

    Instead of armor installed on the first wall, an alternative approach is stand-off termination of runaways on solid particulates injected into the chamber. Given accurate prediction of the runaway beam position, a cloud of particulates could be fired into the path of the beam to scatter and/or absorb the runaway electrons prior to final impact. Such a stand-off runaway termination scheme is similar in principle to the dust shield concept previously considered for steady-state power exhaust at the divertor \cite{Tang-2010-JFusErg}. However, the runaway termination scheme has the advantage that steady-state recycling of solid particulates is not an engineering concern, since runaway termination is a one-off event. Compared to an armor or sacrificial-limiter-based scheme, stand-off runaway termination by solid particulates does not require periodic replacement of armor or limiter components.

    Tungsten particulates are a natural choice for the stand-off runaway termination scheme. Tungsten particulates offer several advantages, including: (1) high atomic number ($Z\mathbin{=}74$) and thus large stopping power and pitch-angle scattering capability; (2) high melting point and cohesive energy allowing significant runaway energy deposition before particulate ablation/melting; and (3) in all-tungsten devices, including ITER and SPARC, deposition of ablated material will not introduce impurities to plasma-facing metal surfaces. Crucially, pitch-angle scattering from tungsten particulates would greatly broaden the distribution of runaway impacts on the first wall, a potentially favorable consideration that must be resolved by tracking the runaway orbits after strong pitch-angle scattering by the tungsten particulates.

    In this work, we elucidate runaway stopping and pitch-angle scattering by tungsten particulates for runaway energies from 1 to 10 MeV using the MCNP code \cite{MCNP6-3}. Furthermore, we investigate secondary electron and gamma ray generation from runaway collisions with tungsten particulates and the resulting contribution to runaway (re)seeding during the termination process. The simulation results show that tungsten particulates have excellent efficacy for runaway stopping and pitch-angle scattering. Large gamma ray emission fluxes are observed, but secondary electron emission from both primary runaway and secondary gamma ray interactions does not appear to pose a serious issue. Furthermore, the data we report here are necessary for the runaway orbit investigation mentioned above, which will be carried out in a separate study.


    \section{Methodology}
    \label{sec:methods}

    The Monte Carlo N-Particle (MCNP) code is a general-purpose Monte Carlo code for simulation of radiation transport in materials \cite{MCNP6-3}. MCNP simulates electron transport by a condensed history method \cite{Berger-1963-MethCompPhys}, in which an electron moves through a selected path length $\delta s$ and experiences energy loss, angular deflection, and secondary particle generation based on statistical aggregation of many collisions over the path length. Statistical distributions for these quantities are derived from physics models of elastic scattering, ionization, and \brems{} processes. The condensed history approach is necessary at high energies due to the large number of collisions an electron experiences in the slowing-down process, but becomes inaccurate for electron energies below 10 keV and breaks down completely at energies below 1 keV. Thus, for low energies we use a single-event method which treats electron transport by successive individual collisions, including elastic, excitation, ionization, and \brems{} collisions \cite{Hughes-2014-PNST}. MCNP simulates photon transport by a single event method, including: incoherent (Compton) scattering, coherent (Thompson) scattering, photoelectric, and pair production collisions.

\begin{figure}[t]
        \centering
        \includegraphics[width=3in]{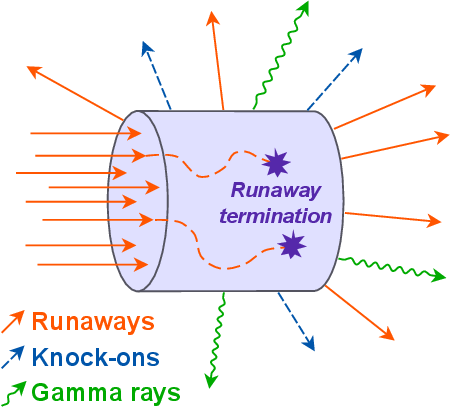}
        \caption{Schematic illustration of MCNP simulation procedure. Runaways are incident into a model tungsten particulate and are scattered in a new direction or terminated. Knock-on electrons (from ionization) and gamma rays (from \brems{}) are generated by runaway collisions and may also exit the particulate.}
        \label{fig:mcnp-setup}
    \end{figure}

    Our MCNP simulation settings are as follows. Simulations are run with the MODE P E card indicating coupled electron-photon transport calculations. The simulation geometry is a solid tungsten cylinder, as shown in \fref{fig:mcnp-setup}, with length ($L$) and diameter ($D$) 1 mm each, chosen to match the average particulate size of the ITER shattered pellet injection systems \cite{Matura-2022-IAEA}. The cylindrical geometry is somewhat arbitrary, and was chosen to support future development of a particulate melting and vaporization model. Runaways are normally incident on a flat circular face of the cylindrical geometry, with a uniform distribution across the exposed surface area. The EL03 data library \cite{Berger-1988-MCTrans,Seltzer-1988-MCTrans} is used for electron transport with the condensed history method, while the EPRDATA14 library \cite{Hughes-2017-ICRS,Cullen-2014-EPICS} is used for photon and low-energy electron transport with the single-event method. The energy boundary between condensed history and single-event methods for electron transport is set at 10 keV, with lower cutoff energies of 100 eV for electrons and 10 eV for photons. Runaways are incident at logarithmically spaced energies of 1.0, 1.6, 2.5, 4.0, 6.3, or 10.0 MeV, and for each energy 1,048,576 ($2^{20}$) histories are simulated.

    We generate two types of simulation outputs. First, we collect energy and angle-resolved fluxes of electrons and photons leaving the pellet by tallying particles of each type which cross a surface placed just beyond the pellet bounding surfaces (the slight offset is used to avoid, e.g., double-counting of source electrons). This type of output is shown schematically in \fref{fig:mcnp-setup}. Energy bins are marked from the lower cutoff energy of the particle type up to 10 MeV with ten logarithmic increments per decade. Angular bins are marked by 200 equal cosine increments from $\xi\equiv\cos\theta\mathbin{=}-1$ to $\xi=1$. Additionally, we use the `FT' input card with the `TAG' option to separate contributions to the total flux exiting the pellet based on the physical process which created the particle, which is necessary to isolate the flux of primary electrons from the flux of secondary electrons from ionization, etc. For electrons, the processes considered include source/primary electrons, photoelectric effect, Compton scattering, pair production, Auger, and knock-on electrons from ionization. For photons, the processes considered include \brems{}, fluorescence, K x-rays, and annihilation photons\footnote{While positrons are not explicitly included in our simulations, MCNP considers non-transported positrons to annihilate and generate a pair of 0.511-MeV photons.}.

    Second, we tally the spatially resolved energy deposition (from all particles and processes) into the pellet using the `TMESH' card over a cylindrical tally mesh, with 128 bins in the radial dimension, 256 bins in the axial ($z$) dimension. This high spatial resolution requires a change to the MCNP source code to obtain accurate energy deposition profiles. Normally, MCNP considers ionization energy loss to be deposited along the substep track length $\delta s$, while energy given to knock-on and Auger electrons is subtracted from the deposited energy at the starting point of the substep. While this is computationally efficient and accurate for most cases, the high spatial resolution used here causes many mesh cells to show unphysical negative energy deposition. For our simulations, we have modified MCNP to remove the energy given to each knock-on and Auger electron at its origin position. This change approximately doubles the runtime of our simulations, so while it is necessary in this work it is neither necessary nor appropriate for more typical cases with sufficiently large mesh cells to contain one or more electron substeps. Finally, the statistical uncertainty of energy deposition for individual mesh cells is often quite extreme due to their small size, particularly near the central axis ($r\rightarrow 0$), so in practice we average tally data over $8\times 8$ sets of mesh cells to obtain meaningful aggregate statistics.


    \section{Results and Discussion}
    \label{sec:results}

    The interactions between runaways and tungsten particulates can be understood from three perspectives: (1) energy loss and deposition, (2) runaway collisional kinematics, and (3) secondary radiation effects. Here, we present the results from each perspective in turn. In each case, the assessment for runaway termination by tungsten particulates appears to be optimistic.

    \subsection{Runaway energy loss and deposition in tungsten particulates}
    \label{subsec:re-energy}

    \begin{figure*}[t]
        \centering
        \begin{subfigure}{5.35in}
            \includegraphics[width=1.75in]{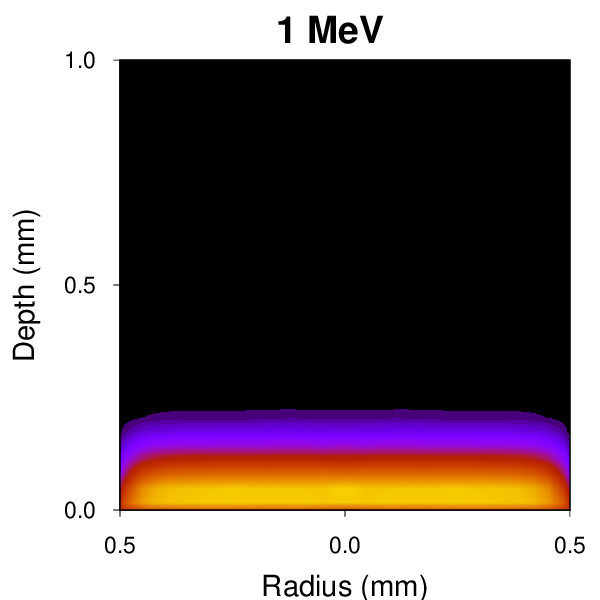}
            \includegraphics[width=1.75in]{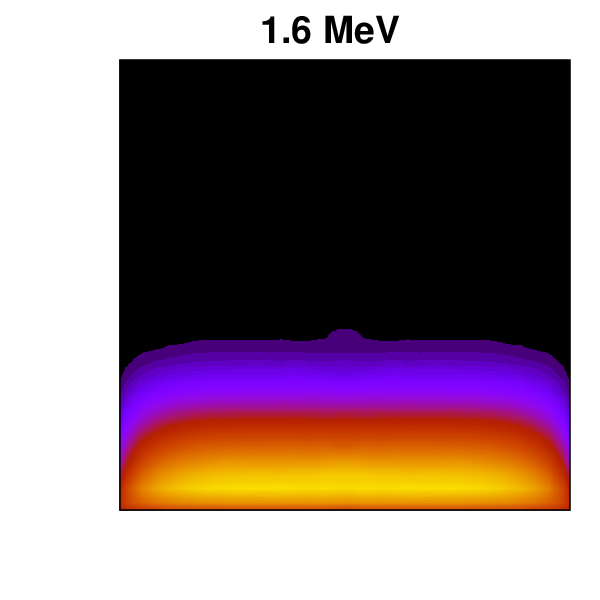}
            \includegraphics[width=1.75in]{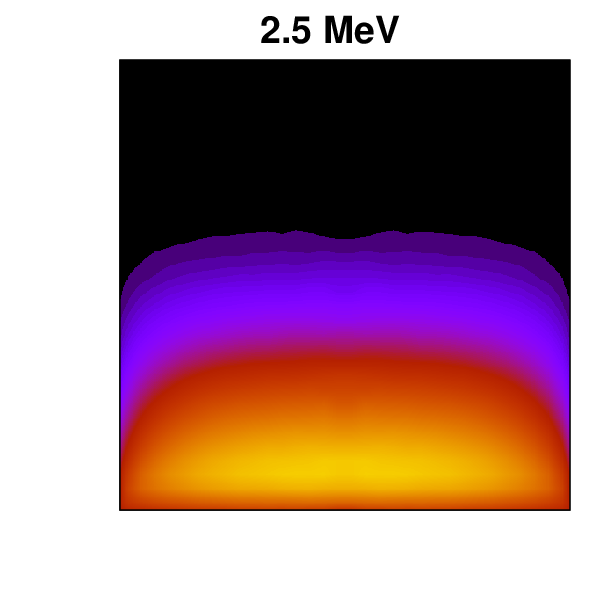}
            \\~\\
            \includegraphics[width=1.75in]{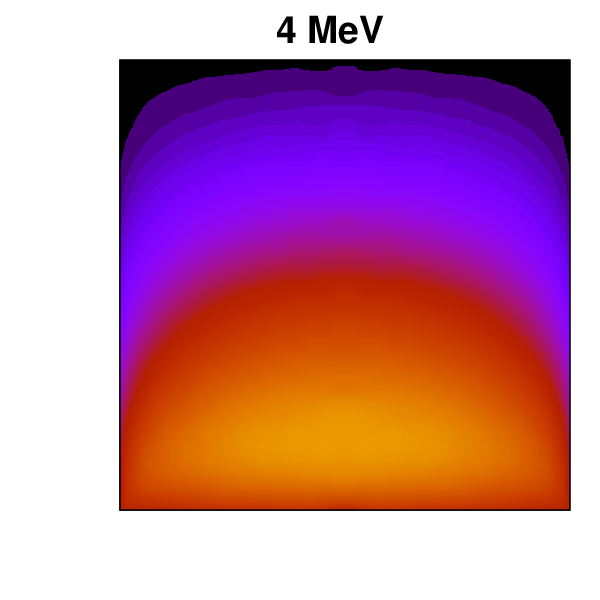}
            \includegraphics[width=1.75in]{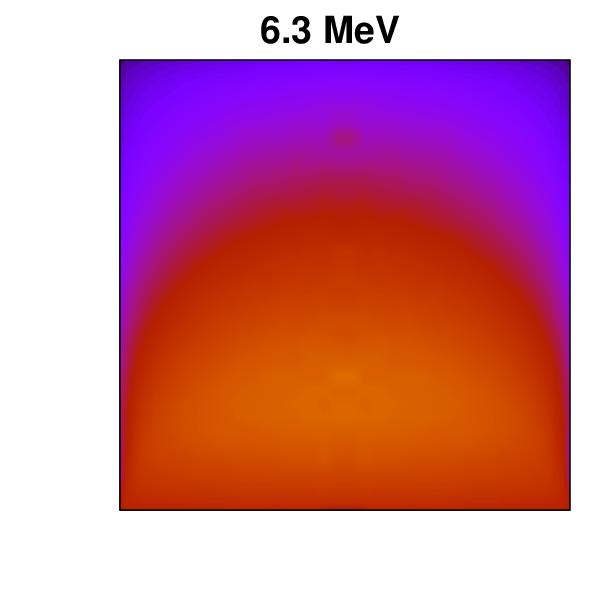}
            \includegraphics[width=1.75in]{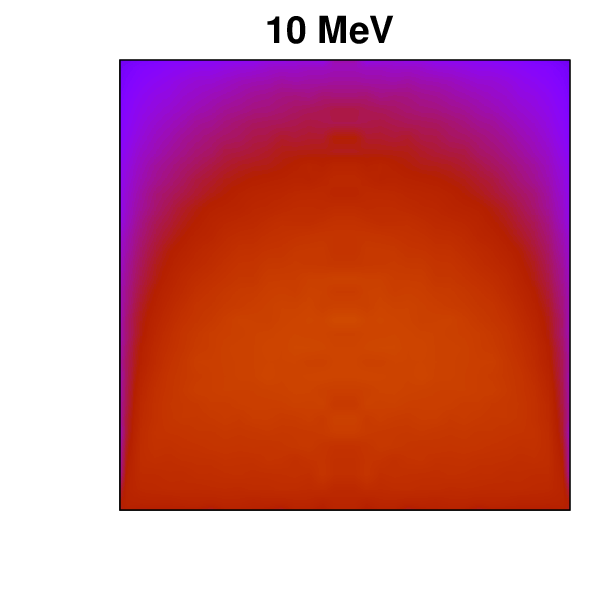}
        \end{subfigure}
        \begin{subfigure}{1in}
            \centering
            \includegraphics[height=3.25in]{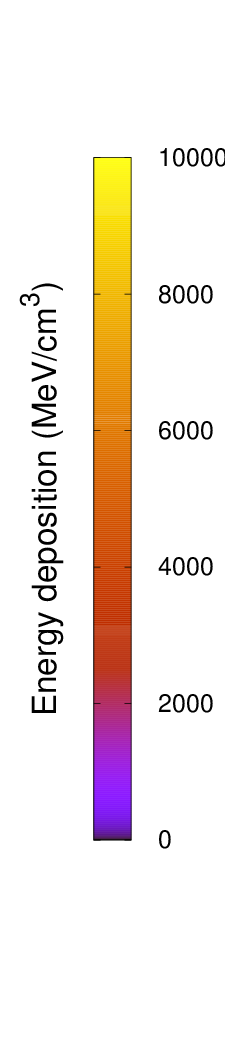}
        \end{subfigure}
        \caption{Volumetric energy deposition profile in tungsten particulates by incident runaways. Energy deposition profiles are plotted along the radial (horizontal) and depth (vertical) axes, with the profiles mirrored across the radial axis to preserve angular symmetry. Runaway incident energies are indicated at the top of each plot.}
        \label{fig:w-ergdep}
    \end{figure*}

    \Fref{fig:w-ergdep} shows volumetric energy deposition profiles for runaways incident on tungsten particulates with incident energies from 1 to 10 MeV. For 6.3 and 10 MeV incident energies, the energy deposition profiles are close to uniform, tapering off towards the far corner (i.e., towards the maximal radial and depth coordinates). This tapering effect reflects the fact that runaways incident near the outer edge of the particulate tend to scatter out from the sides before traversing the full depth. For 4 MeV and lower incident energies, the profiles shift as energy deposition drops to zero beyond some terminal distance. This reflects the fact that lower-energy runaways cannot penetrate through the entire particulate, and will either deposit their full energy before being absorbed inside the material or will scatter backwards.

    Nonuniform energy deposition from incident runaways implies that tungsten particulates will ablate from front to back, rather than by uniform vaporization and expansion. To confirm this, the thermal diffusion time can be computed by
    \begin{equation}
        \label{eq:td-time}
        \tau_\mathrm{td} = \frac{\rho c_\mathrm{p} L^2}{k_\mathrm{c}}
    \end{equation}
    where density $\rho = 19.25~\mathrm{g~cm^{-3}}$, specific heat capacity $c_\mathrm{p} = 0.134~\mathrm{J~(g~K)^{-1}}$, particulate depth $L = 0.1~\mathrm{cm}$, and thermal conductivity $k_\mathrm{c} = 173~\mathrm{W~(m~K)^{-1}}$ yield a thermal diffusion time $\tau_\mathrm{td} = 16.5~\mathrm{ms}$. As shown below, this timescale is much greater than those for particulate melting or vaporization, thus the ablation will proceed front-to-back rather than by fast heat conduction and uniform vaporization. This means that in the event of particulate melting or vaporization, the size of the particulate may change with time, potentially affecting the termination efficacy (although effects such as vapor shielding could mitigate this). Additionally, since a smaller particulate would absorb less energy per incident runaway the particulate may survive longer under the runaway flux, which could also mitigate the effect on termination efficacy.

    \begin{table*}[tb]
        \fontsize{8pt}{10pt}\selectfont  
        \caption{Key statistics for runaway energy loss and deposition in tungsten particulates. $E_0$: incident runaway energy. $E_\mathrm{out}$: average runaway exit energy. $E_\mathrm{see}$: average energy lost to secondary electron emission. $E_\gamma$: average energy lost to gamma radiation. $E_\mathrm{rad} = E_\mathrm{see} + E_\gamma$: total average radiative energy loss. $E_\mathrm{dep}$: average energy deposited into the tungsten particulate. $\Delta\epsilon_\mathrm{max}$: Peak volumetric energy deposition. $\tau_\mathrm{erg}$: time to absorb 99\% of energy from runaways with initial energy $E_0$. $\tau_\mathrm{m}$ and $\tau_\mathrm{v}$: estimated melting and vaporization times under ablation by runaways incident with uniform energy $E_0$. Statistical uncertainty in the last one or two digits is given in parentheses for each reported value.}
        \label{tab:re-energy}
        \lineup
        \begin{tabular}{@{}lllllllll}
            \br
            $E_0$ (MeV) & $E_\mathrm{out}$ (MeV) & $E_\mathrm{see}$ (MeV) & $E_\gamma$ (MeV) & $E_\mathrm{dep}$ (MeV) & $\Delta\epsilon_\mathrm{max}$ (MeV cm$^{-3}$) & $\tau_\mathrm{erg}$ ($\mu$s) & $\tau_\mathrm{m}$ ($\mu$s) & $\tau_\mathrm{v}$ ($\mu$s) \\
            \mr
            \01.0 & 0.38(4)  & 0.0021(2) & 0.026(3) & 0.5963(2) & $1.2(2) \0\0\times 10^4$ & 4.48(2)  & \09.6(14) &  150.76(4)   \\
            \01.6 & 0.61(7)  & 0.0041(4) & 0.065(7) & 0.9200(2) & $1.16(15)   \times 10^4$ & 4.53(4)  & \09.9(12) & \097.72(3)   \\
            \02.5 & 1.03(11) & 0.0079(8) & 0.15(2)  & 1.3120(3) & $1.2(6) \0\0\times 10^4$ & 4.77(6)  & \09.(4)   & \068.53(2)   \\
            \04.0 & 1.9(2)   & 0.016(2)  & 0.35(4)  & 1.7217(5) & $1.1(4) \0\0\times 10^4$ & 5.42(11) &  10.(4)   & \052.220(14) \\
            \06.3 & 3.6(4)   & 0.034(3)  & 0.72(8)  & 1.9308(6) & $1.1(7) \0\0\times 10^4$ & 6.7(2)   &  10.(6)   & \046.565(14) \\
            10.0 & 6.5(7)   & 0.072(6)  & 1.37(15) & 2.0102(8) & $1.0(4) \0\0\times 10^4$ & 8.4(4)   &  11.(4)   & \044.72(2)   \\
            \br
        \end{tabular}
    \end{table*}

    \Tref{tab:re-energy} gives the major statistical quantities relating to runaway energy loss and deposition in tungsten particulates. The first key quantity is the average exit energy of a runaway electron, $E_\mathrm{out}$, from which the average energy lost per incident runaway is $E_\mathrm{loss} = E_0 - E_\mathrm{out}$. Runaway energy losses are classified as radiation losses ($E_\mathrm{rad}$), to secondary electron emission ($E_\mathrm{see}$) and to gamma radiation ($E_\gamma$), and as energy deposited into the tungsten itself ($E_\mathrm{dep}$). The majority of energy lost at all incident energies is deposited into the tungsten particulate. However, while the absolute energy deposition increases with increasing incident energy, the relative magnitude (i.e., as a fraction of $E_0$) decreases with incident energy, from 60\% deposition at 1 MeV down to only 20\% deposition at 10 MeV. Radiation losses are the minority component at all energies, but increase in both absolute and relative magnitude with increasing incident energy which partially offsets the relative decrease in energy deposition.

    Given the average energy loss, we can immediately assess the runaway termination in energy terms,
    \begin{equation}
        \label{eq:term-erg}
        F_\mathrm{term} = 1 - \left(1 - f_\mathrm{area}f_\mathrm{erg}\right)^{t / \tau_\mathrm{re}}
    \end{equation}
    where $f_\mathrm{area}\approx 20\%$ is the fraction of the runaway beam cross-sectional area intersected by tungsten particulates, $f_\mathrm{erg} = E_\mathrm{loss} / E_0$ is the average fraction of incident energy lost by a colliding runaway, and $\tau_\mathrm{re} = 2\pi R_0 / c = 0.13~\mathrm{\mu s}$ is the runaway orbital period in ITER ($R_0 = 6.2~\mathrm{m}$), indicating that runaways orbiting the tokamak will have multiple chances to collide with a particulate, or may strike a particulate multiple times. We find that the time to remove 99\% of runaway energy, denoted as $\tau_\mathrm{erg}$ in \tref{tab:re-energy}, is less than 10 $\mu$s for all simulated incident energies. In other words, 99\% of runaway energy will be removed in fewer than $10^2$ runaway toroidal transits. Therefore, tungsten particulates are effective for removing nearly all runaway energy in a very short time.

    As an aside, the total volume of tungsten needed to cover $f_\mathrm{area}\approx 20\%$ may be estimated by
    \begin{equation}
        \label{eq:w-volume}
        V_\mathrm{tot} = f_\mathrm{area}\frac{I_\mathrm{re}}{j_\mathrm{re}}\frac{V_\mathrm{par}}{A_\mathrm{par}}
    \end{equation}
    where the ratio of runaway current to runaway current density, $I_\mathrm{re} / j_\mathrm{re}$, gives the cross-sectional area of the runaway beam. For typical values $I_\mathrm{re} = 5~\mathrm{MA}$ and $j_\mathrm{re} = 1~\mathrm{MA~m^{-2}}$ gives a cross-sectional area of 5 m$^2$. Multiplying by $f_\mathrm{area}$ yields a total particulate cross section of 1 m$^2$. Dividing this by the particulate cross-sectional area, $A_\mathrm{par} = \pi D^2/4 = 7.85\times 10^{-3}~\mathrm{cm^2}$, gives the total number of particulates to inject, and multiplying this by the particulate volume, $V_\mathrm{par} = A_\mathrm{par}L = 7.85\times 10^{-4}~\mathrm{cm^3}$, gives the total volume. For the given conditions, we find $V_\mathrm{tot} = 10^3~\mathrm{cm^3}$ of tungsten is required to achieve $f_\mathrm{area} = 20\%$, which is 19.25 kg after multiplying by density. This highlights the advantage of stand-off termination by particulates rather than terminating runaways at the wall, since to achieve the same volume of tungsten at a runaway final impact spot size of $\sim$100 cm$^2$ would require a rather impractical armor thickness of 10 cm. Note that $f_\mathrm{area}$, along with the particulate size, may be considered optimizable parameters to minimize the required volume of material, as discussed later.

    To assess if the energy removal time, $\tau_\mathrm{erg}$, is short enough, we also estimate the melting and vaporization times for the tungsten particulates. The characteristic time for onset of melting is
    \begin{equation}
        \label{eq:tau-melt}
        \tau_\mathrm{m} = \frac{e}{A_\mathrm{par} j_\mathrm{re}}\times\frac{\Delta H_\mathrm{m}^\circ N}{\Delta\epsilon_\mathrm{max}}
    \end{equation}
    where $e$ is the electronic charge, $j_\mathrm{re}\sim 1~\mathrm{MA~m^{-2}}$ is the runaway current density, $\Delta H_\mathrm{m}^\circ = 1.594~\mathrm{eV}$ is the per-atom enthalpy of melting, $N = 6.306\times 10^{22}~\mathrm{cm^{-3}}$ is the atomic density of tungsten, $A_\mathrm{par}$ is the particulate cross-sectional area given previously, and $\Delta\epsilon_\mathrm{max}$ is the \textit{peak} volumetric energy deposition (in MeV cm$^{-3}$), given in \tref{tab:re-energy}. Physically, since the peak rate of energy deposition is the product of runaway-particulate collision frequency ($A_\mathrm{par} j_\mathrm{re} / e$) and peak volumetric energy deposition per runaway ($\Delta\epsilon_\mathrm{max}$), dividing the volumetric melting enthalpy ($\Delta H_\mathrm{m}^\circ N$) by the peak rate of energy deposition yields the time before onset of melting. Using the peak energy deposition yields the earliest possible onset time for particulate melting, providing a conservative lower bound on the particulate lifetime as shown in \tref{tab:re-energy}. Inspecting \tref{tab:re-energy} shows that for all simulated runaway energies, the energy removal time is less than the onset-of-melting time, i.e., $\tau_\mathrm{erg} < \tau_\mathrm{m}$. Thus, tungsten particulates appear capable of removing nearly all runaway energy before any tungsten can melt and splatter on the first wall.

    For completeness, we also estimate the particulate lifetime before total vaporization,
    \begin{equation}
        \label{eq:tau-vapor}
        \tau_\mathrm{v} = \frac{e}{A_\mathrm{par} j_\mathrm{re}}\times\frac{E_\mathrm{coh} NV_\mathrm{par}}{E_\mathrm{dep}}
    \end{equation}
    where $E_\mathrm{coh} = 8.90~\mathrm{eV}$ is the cohesive energy of tungsten, $E_\mathrm{dep}$ is the energy deposition from \tref{tab:re-energy}, and $V_\mathrm{par}$ is the particulate volume given previously. Physically, this is similar to \eref{eq:tau-melt}, giving the ratio of the energy required for complete vaporization ($E_\mathrm{coh} NV_\mathrm{par}$) over the average energy deposition rate per runaway ($E_\mathrm{dep} \times A_\mathrm{par} j_\mathrm{re} / e$). This provides an approximate upper bound for runaway termination by tungsten particulates. In the event that runaways are not completely terminated before the onset of melting, the particulates should at least remain in a condensed state until time $\tau_\mathrm{v}$ and thus remain effective. Note that this only a rough estimate, neglecting important material effects such as vapor shielding which can also contribute to runaway termination.

    While the calculations in \tref{tab:re-energy} give a favorable picture for runaway termination from an energy perspective, we note that the large uncertainty in $\tau_\mathrm{m}$ as well as the possibility for larger values of $j_\mathrm{re}$ do raise the possibility that particulate melting could occur nevertheless. In this case, complete vaporization of the particulates may be preferable, since the broadly uniform deposition of vaporized tungsten on the first wall is preferable to droplets splashing on the wall, since the latter could pose significant plasma-material interactions problems during later operation. In this case, an optimal particulate size would be desired such that $\tau_\mathrm{erg} \gtrsim \tau_\mathrm{v}$ (counting on the vaporized tungsten to finish the runaway termination process). Such size optimization may also be desirable to reduce the mass of tungsten required (i.e., reducing $f_\mathrm{area}$). Investigations to determine an optimal particulate size for both complete vaporization and runaway termination are therefore a potential avenue for future studies.

    \subsection{Runaway collisional kinematics with tungsten particulates}
    \label{subsec:re-kinetics}

    While the energy removal time calculated above is a convincing metric for the runaway termination efficacy of tungsten particulates, it may not give the full picture. Runaways which lose energy on collision with a tungsten particulate may be accelerated by the toroidal electric field and regain some energy, complicating the picture. Runaway orbit simulations which include this acceleration can provide further confidence, but are beyond the scope of the present work and will be carried out in a separate study using the MCNP data produced in current study. Therefore, here we present runaway collisional data to qualitatively assess the ability of tungsten particulates to reduce the runaway population from a collisional kinematics perspective. In this perspective, tungsten particulates terminate colliding runaways by (1) pitch-angle scattering of runaways into magnetically trapped orbits or directional reversal (backscattering), leading to rapid deceleration, or by (2) directly absorbing runaways into the particulate.

    \begin{figure*}[t]
        \centering
        \begin{subfigure}{5.35in}
            \includegraphics[width=1.75in]{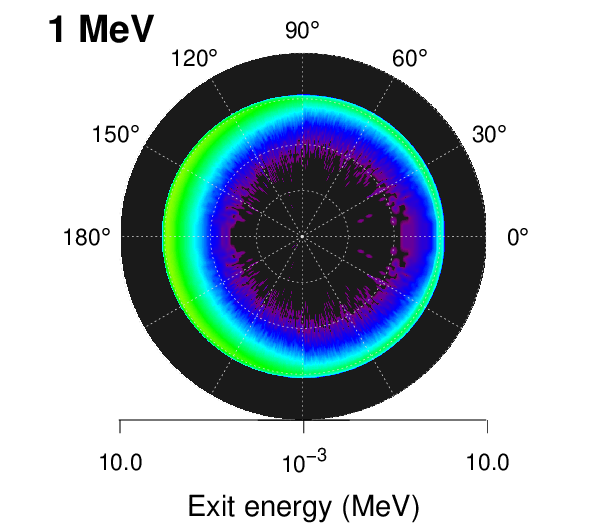}
            \includegraphics[width=1.75in]{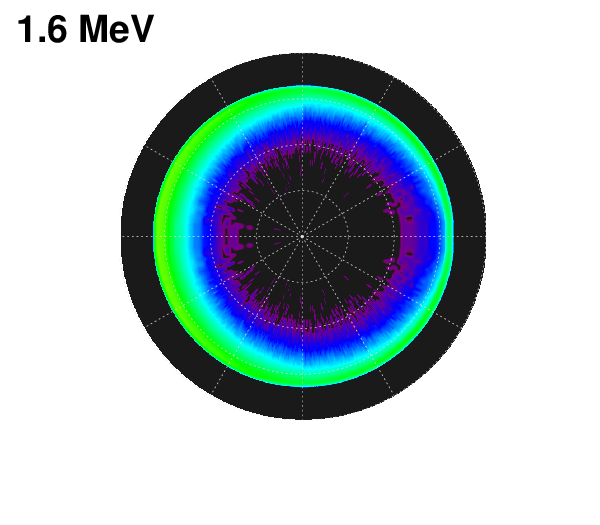}
            \includegraphics[width=1.75in]{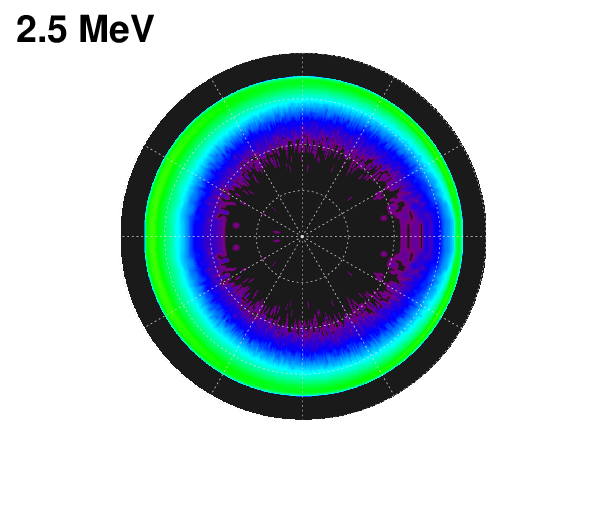}
            \\~\\
            \includegraphics[width=1.75in]{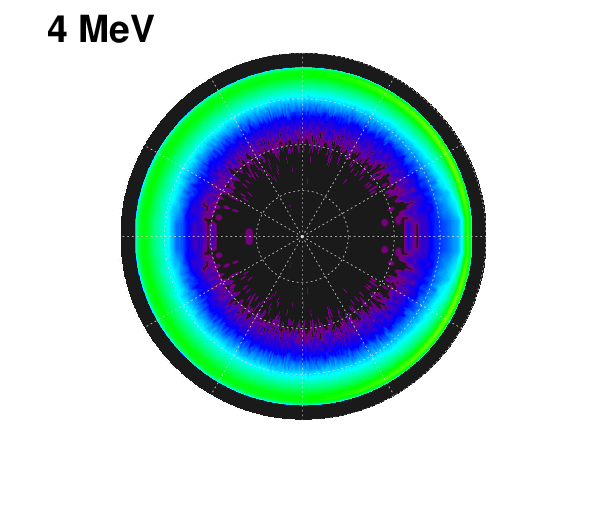}
            \includegraphics[width=1.75in]{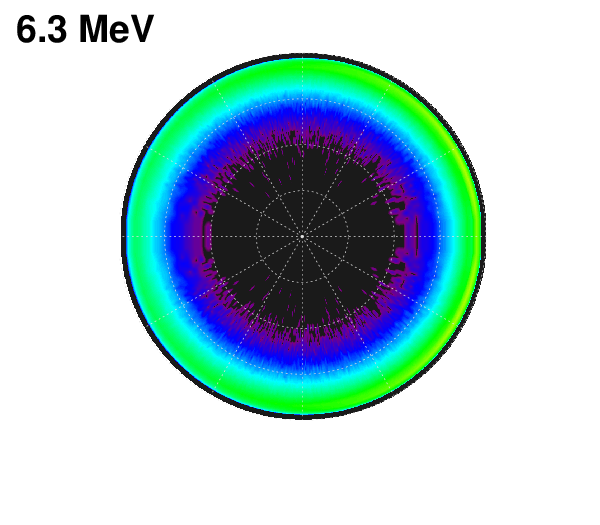}
            \includegraphics[width=1.75in]{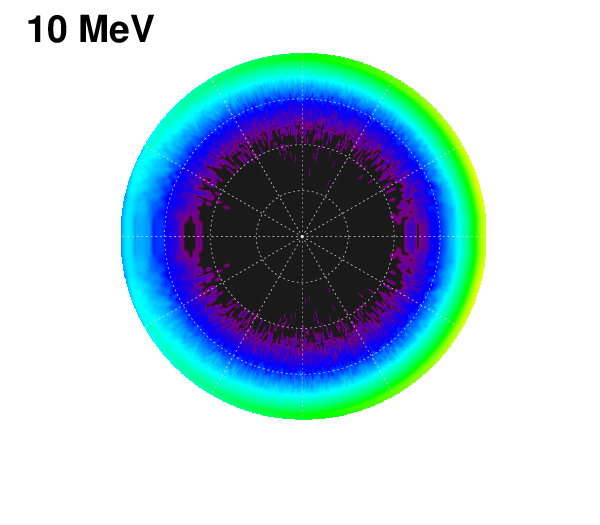}
        \end{subfigure}
        \begin{subfigure}{1in}
            \centering
            \includegraphics[height=3.25in]{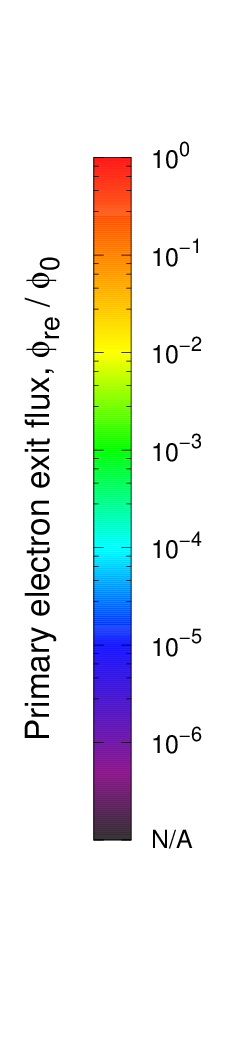}
        \end{subfigure}
        \caption{Energy and angle-resolved fluxes of primary runaways exiting a tungsten particulate. The radial axis indicates the energy with which primary electrons exit the material and the angular axis indicates the exit direction, where $0^\circ$ is the incident direction. Exiting fluxes are normalized against the incident runaway flux with values corresponding to the color bar at right. Runaway incident energies are indicated at the top-right of each plot.}
        \label{fig:w-ergcos}
    \end{figure*}

    \Fref{fig:w-ergcos} shows energy and angular-resolved fluxes of runaways exiting from simulated particulates. Strong pitch-angle scattering is visible as the spread of data away from the $0^\circ$ angular mark. While significant scattering through the full angular range occurs for all incident runaway energies, there is a slight tendency toward forward scattering at 6.3 and 10 MeV, whereas for 4 MeV and lower energies backscattering is more prevalent. The shift towards backscattering arises because the lower-energy runaways are unable to penetrate through the tungsten particulate (see corresponding energy deposition plots in \fref{fig:w-ergdep}). In the same figure, the runaway energy attenutation is visible as the spread of the distribution along the energy (radial) axis below the incident energy. The energy distribution is not strongly correlated with the angular distribution in any case. It is visually evident that while most exiting runaways lose up to a decade (90\%) of the incident energy, these remain relativistic electrons (i.e., $E \ge 100~\mathrm{keV}$).

    \begin{table*}[tb]
        \caption{Key statistics for runaway collisional kinematics and termination by tungsten particulates. $E_0$: incident runaway energy. $f_\mathrm{trap}$: fraction of exiting runaways with pitch-angle scattering into the trapped zone, $|\xi| \le \xi_c\equiv \sqrt{2a/R_0}$. $f_\mathrm{back}$: backscattered fraction of runaways, $\xi<-\xi_c$. $f_\mathrm{abs}$: fraction of incident runaways absorbed inside the particulate. $f_\mathrm{term}$: terminated fraction, $f_\mathrm{trap} + f_\mathrm{back} + f_\mathrm{abs}$. $\tau_\mathrm{par}$: time to terminate 99\% of the runaway population by pitch-angle scattering or absorption. Statistical uncertainty in the last one or two digits is given in parentheses for each reported value.}
        \label{tab:re-kinetics}
        \lineup
        \begin{tabular}{@{}llllll}
            \br
            $E_\mathrm{0}$ (MeV) & $f_\mathrm{trap}$ (\%) & $f_\mathrm{back}$ (\%) & $f_\mathrm{abs}$ (\%) & $f_\mathrm{term}$ (\%) & $\tau_\mathrm{par}$ ($\mu$s) \\
            \mr
            \01.0 & 34.58(6) &  15.68(4)   &  48.53(6) & 98.80(9) & 2.718(3) \\
            \01.6 & 38.34(6) &  13.34(4)   &  46.04(6) & 97.72(9) & 2.752(3) \\
            \02.5 & 45.10(7) &  10.37(3)   &  40.44(6) & 95.92(9) & 2.810(3) \\
            \04.0 & 56.28(7) & \06.86(3)   &  29.31(5) & 92.45(9) & 2.927(3) \\
            \06.3 & 67.00(8) & \03.55(2)   &  14.12(3) & 84.68(9) & 3.225(4) \\
            10.0 & 62.38(8) & \01.284(11) & \05.74(2) & 69.40(8) & 4.005(5) \\
            \br
        \end{tabular}
    \end{table*}

    \begin{figure}[t]
        \centering
        \includegraphics[width=3in]{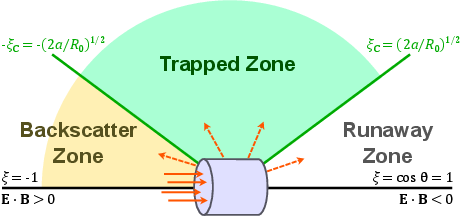}
        \caption{Diagram of pitch-angle scattering zones for runaway electrons. Three possibilities are considered: (1) runaway pass-through zone with scattering cosine $\xi > \xi_\mathrm{c}$, (2) trapped zone with scattering cosine $|\xi| \le \xi_\mathrm{c}$, and (3) backscattering zone with scattering cosine $\xi < -\xi_\mathrm{c}$. Trapping and backscattering, along with absorption in the particulate, contribute to runaway termination. The direction of runaway acceleration by the electric field $\mathbf{E}$ is parallel to the magnetic field $\mathbf{B}$ where $\mathbf{E\cdot B} < 0$.}
        \label{fig:scat-zones}
    \end{figure}

    \Tref{tab:re-kinetics} gives the major statistical quantities relating to runaway collisional kinematics. On first inspection, the two key quantities are the fraction of runaways which experience strong pitch-angle scattering ($f_\mathrm{scat}$) or which are absorbed into the particulate ($f_\mathrm{abs}$). We define strong pitch-angle scattering in terms of the scattering cosine $\xi$ as follows: first, we consider a critical value $\xi_\mathrm{c} = \sqrt{2a / R_0}$, where $a$ is the tokamak minor radius and $R_0$ the major radius. For ITER ($a = 2.0~\mathrm{m}, R_0 = 6.2~\mathrm{m}$) this gives $\xi_\mathrm{c} = 0.8$ or a scattering angle of 38.8$^\circ$. This critical value is the boundary between runaway passing-through and trapped orbit zones, illustrated in \fref{fig:scat-zones}. Scattered runaways with $\xi > \xi_\mathrm{c}$ are considered to pass through and continue along a runaway orbital trajectory. Scattered runaways with $-\xi_\mathrm{c} < \xi < \xi_\mathrm{c}$ will enter trapped orbits and will lose energy due to enhanced synchrotron radiation, and thus are considered terminated ($f_\mathrm{trap}$). Finally, scattered runaways with $\xi < -\xi_\mathrm{c}$ are backscattered and will be decelerated by the parallel electric field before falling into the trapped zone, and thus are also considered terminated ($f_\mathrm{back}$). Therefore, we define the fraction of runaways terminated by pitch-angle scattering as $f_\mathrm{scat} = f_\mathrm{trap} + f_\mathrm{back}$.

    Thus, the fraction of runaway population which has been eliminated by time $t$ may be estimated from
    \begin{equation}
        \label{eq:term-par}
        F_\mathrm{term} = 1 - (1 - f_\mathrm{area} f_\mathrm{term})^{t/\tau_\mathrm{re}}
    \end{equation}
    following the definitions given with \eref{eq:term-erg} and with $f_\mathrm{term} = f_\mathrm{scat} + f_\mathrm{abs}$. The times required to conclusively eliminate 99\% of the runaway population, $\tau_\mathrm{par}$, at each runaway energy are given in the latter column of \tref{tab:re-kinetics}. Comparing with the melting and vaporization times in \tref{tab:re-energy}, the values of $\tau_\mathrm{par}$ for collisional termination are less than the melting and vaporization times at all energies. Therefore, from the collisional kinematics perspective tungsten particulates appear capable of eliminating nearly all runaways from the population before any tungsten can melt and splatter on the first wall.

    \subsection{Secondary radiation emission from tungsten particulates}
    \label{subsec:radiation}

    The final important phenomenon arising from runaway collisions with tungsten particulates is the emission of secondary radiation. This comes in two flavors: secondary electron emission may produce additional runaways in the form of relativistic secondary electrons, while gamma radiation may also produce additional runaways if the gamma rays interact with other tungsten particulates. We consider both relativistic electrons and gamma rays to have $E\ge 100$ keV for the purposes of this work. Interested readers may find energy and angular-resolved distributions for secondary electron and photon fluxes in the \siname{}.

    \begin{table*}[tb]
        \caption{Key statistics for secondary radiation emitted from tungsten particulates. $E_0$: incident runaway energy. $N_\mathrm{se}, N_\gamma$: number of secondary electrons and gamma rays emitted, respectively, reported per incident     runaway ($N_\mathrm{re}$). $\langle E_\mathrm{se}\rangle, \langle E_\gamma\rangle$: average emitted energy of secondary     electrons and gamma rays, respectively. Statistical uncertainty in the last one or two digits is given in parentheses for each reported value.}
        \label{tab:re-radiation}
        \lineup
        \begin{tabular}{@{}lllll}
            \br
            $E_\mathrm{0}$ (MeV) & $N_\mathrm{se} / N_\mathrm{re}$ & $\langle E_\mathrm{se}\rangle$ (MeV) & $N_\gamma / N_\mathrm{re}$ & $\langle E_\gamma\rangle$ (MeV) \\
            \mr
            \01.0 & 0.0081(9)  & 0.26(4)  & 0.0755(3)  & 0.34(4)  \\
            \01.6 & 0.0128(10) & 0.32(4)  & 0.1563(4)  & 0.42(5)  \\
            \02.5 & 0.0191(10) & 0.41(5)  & 0.2890(5)  & 0.53(6)  \\
            \04.0 & 0.0303(11) & 0.53(6)  & 0.4932(7)  & 0.71(8)  \\
            \06.3 & 0.0456(12) & 0.74(7)  & 0.7201(8)  & 1.00(11) \\
            10.0 & 0.0631(12) & 1.14(11) & 0.9370(10) & 1.5(2)   \\
            \br
        \end{tabular}
    \end{table*}

    \Tref{tab:re-radiation} gives the major statistical quantities related to secondary radiation emission from tungsten particulates. Generally, gamma radiation dominates over secondary electron emission with an order of magnitude greater emission rate and modestly greater average energy per emitted particle. However, both types of secondary radiation carry away only a fraction of the incident runaway energy, which means that even if secondary radiation leads to runaway reseeding, the overall energy of the runaway population is still reduced.

    \begin{figure}[t]
        \centering
        \includegraphics[width=3in]{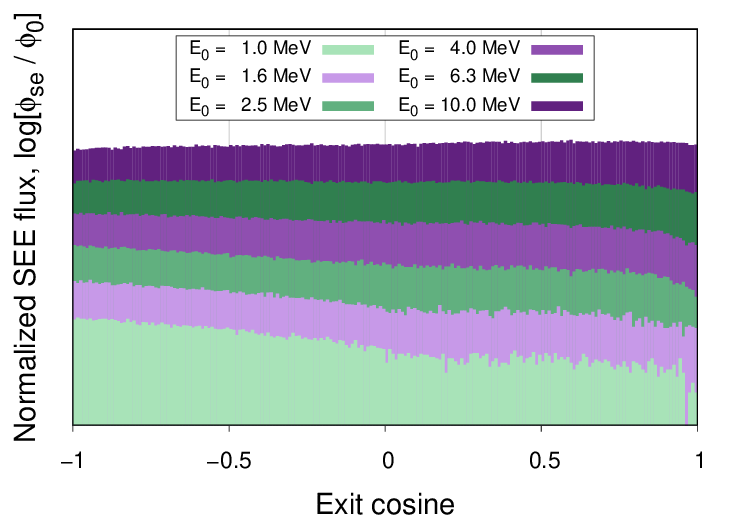}
        \caption{Exit cosine distributions for secondary electron emission (SEE) from tungsten particulates. The incident runaway energy for each case, ranging from 1 to 10 MeV, is indicated in the figure legend. The vertical axis has arbitrary units, and distributions for each runaway energy are offset vertically for visual clarity.}
        \label{fig:see-cos}
    \end{figure}

    The angular distributions of relativistic secondary electrons ($E\ge 100$ keV) emitted from tungsten particulates are shown in \fref{fig:see-cos}, in terms of the exit cosine $\xi = \cos\theta$. At every runaway energy, the angular distribution is extremely broad, with a rather uniform distribution for 10 MeV runaways and a gradual shift away from forward emission and toward backward emission as runaway energy decreases. This implies that the vast majority of secondary electrons will not fall into the passing cone and be accelerated by the parallel electric field. Even for those born inside the passing cone, their pitch is usually large enough that synchrotron radiation can overwhelm the parallel electric field acceleration, which is the primary cause for the runaway vortex flow in momentum space~\cite{Guo-2017-PPCF}. In other words, the broad pitch distribution of the secondary electrons will lead very few of them to be promptly accelerated as new runaways. Combined with the low rates of secondary electron emission seen in \tref{tab:re-radiation}, we conclude that secondary electron emission from tungsten particulates will have a minimal impact on overall termination efficacy.


    \begin{table*}[tb]
        \caption{Key statistics for gamma ray collisions with tungsten particulates. $E_0$: incident gamma ray energy. $f_\mathrm{abs}$: fraction of incident gamma rays absorbed inside the particulate. $N'_\gamma, N_\mathrm{se}$: number of secondary gamma rays and relativistic electrons emitted, respectively, reported per incident gamma ray ($N_\gamma$). $\langle E_{\gamma'}\rangle, \langle E_\mathrm{se}\rangle$: average emitted energy of secondary gamma rays and relativistic electrons, respectively. Statistical uncertainty in the last one or two digits is given in parentheses for each reported value.}
        \label{tab:gamma-stats}
        \lineup
        \begin{tabular}{@{}llllll}
            \br
            $E_0$ (MeV) & $f_\mathrm{abs}$ (\%) & $N'_\gamma / N_\gamma$ & $\langle E_{\gamma'}\rangle$ (MeV) & $N_\mathrm{se} / N_\gamma$ & $\langle E_\mathrm{se}\rangle$ (MeV) \\
            \mr
            \01.0 & 3.172(7) & 0.0040(5) & 0.39(3) & 0.01319(11) & 0.55(6)  \\
            \01.6 & 1.812(5) & 0.0133(6) & 0.48(4) & 0.01736(13) & 0.84(9)  \\
            \02.5 & 2.164(5) & 0.0336(6) & 0.52(5) & 0.0259(2)   & 1.22(14) \\
            \04.0 & 3.276(6) & 0.0574(6) & 0.55(5) & 0.0444(2)   & 1.8(2)   \\
            \06.3 & 4.734(7) & 0.0770(7) & 0.62(5) & 0.0745(3)   & 2.7(3)   \\
            10.0 & 6.481(8) & 0.0935(7) & 0.81(7) & 0.1156(3)   & 4.3(5)   \\
            \br
        \end{tabular}
    \end{table*}

    Gamma rays emitted from one tungsten particulate may produce relativistic secondary electrons through collisions with other nearby particulates. These secondary electrons result chiefly from \brems{} interactions, with minority contributions from fluorescence and pair production collisions. We carried out additional simulations to assess relativistic electron production by gamma radiation, with the same setup as in \fref{fig:mcnp-setup} but with photons instead of electrons as the source particles. \Tref{tab:gamma-stats} summarizes the key results from these simulations. Note that both the emission rate and emitted energy for secondary gamma rays are fairly low for all incident energies, with most of these secondary gamma rays resulting from \brems{} interactions of secondary electrons, with minority contributions from fluorescence and positron annihilation. Even for 10 MeV incident gamma rays, the average energy of a secondary gamma ray is less than 1 MeV. Since the production of relativistic secondary radiation by 1 MeV gamma rays is extremely low, secondary gamma radiation is not a significant consideration for runaway reseeding. In other words, gamma radiation chains leading to secondary electron emission (e.g., $\gamma \rightarrow \gamma \rightarrow \mathrm{e^-}$) are negligible as a source of relativistic secondary electrons.

    The rates of relativistic electron emission by gamma ray collisions are similar to the rates of gamma radiation, with the emitted secondary electron flux ranging from 1\% to 12\% of the incident gamma ray flux. However, the emitted energies of secondary electrons are of the order $\sim$50\% of the incident gamma ray energy. These energies are much greater than not only the secondary gamma radiation energies but also the secondary radiation energies from runaway collisions with the particulates. Secondary electron emission at these energies could contribute significantly to the runaway population.

    \begin{figure}[t]
        \centering
        \begin{subcaptiongroup*}
            \includegraphics[width=\columnwidth]{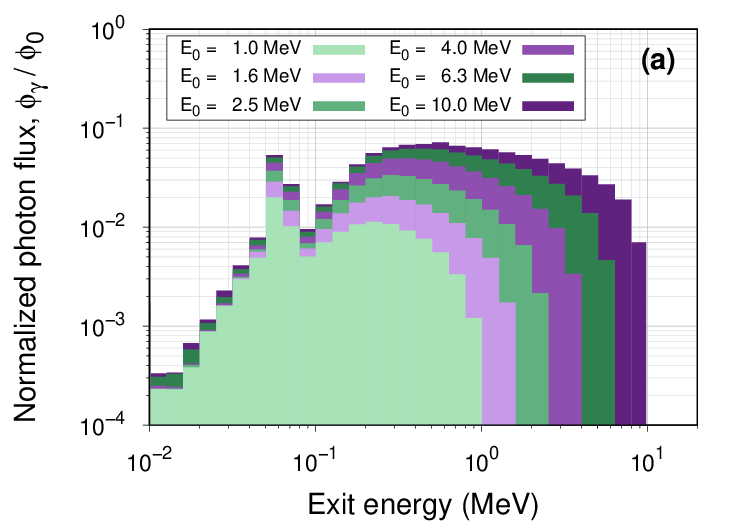}
            \phantomcaption\label{subfig:gamma-erg}
            \includegraphics[width=\columnwidth]{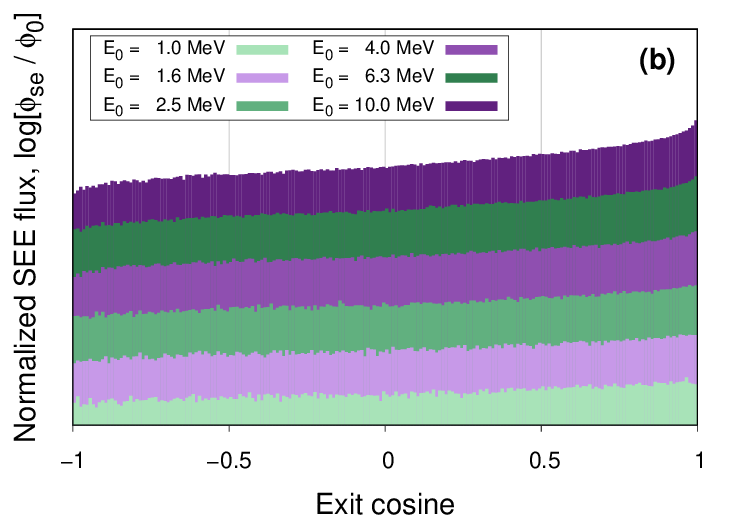}
            \phantomcaption\label{subfig:gamma-see-cos}
        \end{subcaptiongroup*}
        \caption{(\subref{subfig:gamma-erg}) Energy distribution of gamma radiation by runaway collisions with tungsten particulates, for incident runaway energies from 1 to 10 MeV. Gamma ray fluxes are normalized to the incident runaway fluxes in each case. (\subref{subfig:gamma-see-cos}) Exit cosine distributions for secondary electron emission (SEE) by gamma ray interactions with tungsten particulates, for incident gamma ray energies from 1 to 10 MeV. The vertical axis has arbitrary units, and distributions for each gamma ray energy are offset vertically for visual clarity.}
        \label{fig:gamma-info}
    \end{figure}

    In light of these large secondary electron energies from gamma ray collisions, we give two reasons for optimism. First, we note that the energy distribution of gamma radiation from tungsten particulates has a broad spread and is concentrated away from the highest gamma ray energies. \Fref{subfig:gamma-erg} shows the energy distributions for gamma radiation from tungsten particulates induced by runaways from 1 to 10 MeV. At every incident runaway energy, the majority of gamma radiation has energies less than 1 MeV, and the distribution falls off sharply (i.e., by an order of magnitude) towards the highest energy of each distribution. Thus, only a minority of gamma rays can produce relativistic secondary electrons. In the end, the majority of emitted gamma rays will produce a very small number of relativistic electrons (1 per 100 gamma rays or fewer), and most of these with energies well below 1 MeV.

    Second, as for secondary electron emission by runaway collisions, the angular distribution for secondary electron emission by gamma ray collisions is also quite broad. These distributions are shown in \fref{subfig:gamma-see-cos} for each incident gamma ray energy. Since these distribution are spread out over the whole angular range, relatively few emitted relativistic electrons will have the narrow pitch that can be re-accelerated by the parallel electric field. Note that while these distributions do show a preference for forward emission at higher incident gamma ray energies, the forward direction is relative to the incident gamma ray trajectory, which is not in general aligned with magnetic field lines (see the \siname{} for details of gamma ray angular distributions). In other words, an emitted secondary electron is rarely aligned with the magnetic field and thus rarely experiences parallel electric field re-acceleration.

    We note that beside gamma ray interactions with tungsten particulates, two other potential runaway reseeding mechanisms could be considered: Compton scattering of cold plasma electrons and interactions with the first wall. In the former case, the maximum cross section for Compton scattering is the Thomson cross section, $\sigma_\mathrm{T} = 6.65\times 10^{-25}~\mathrm{cm^2}$. For an electron density of $n\sim 4\times 10^{20}~\mathrm{m^{-3}}$ \cite{MartinSolis-2015-PhysPlas} this gives a mean free path on the order of $10^9~\mathrm{m}$, far exceeding the dimensions of any possible terrestrial fusion device. Therefore, Compton scattering is not a significant runaway seeding mechanism. In the latter case, we have carried out additional simulations of gamma ray incidence at the tungsten first wall. The resulting secondary radiation fluxes in these simulations were three orders of magnitude less than those resulting from gamma ray collisions with tungsten particulates. This is because secondary radiation can only emerge from the first wall on a near-backwards trajectory (i.e., $\xi \approx -1$), whereas secondary radiation may emerge from a small particulate in any direction. Therefore, gamma ray interactions with the first wall are also not a significant runaway seeding mechanism.

    To summarize, while runaway collisions with tungsten particulates can lead to relativistic secondary electron emission and gamma radiation, further MCNP simulations show that neither of these secondary radiation types pose a serious challenge to runaway termination efficacy. On one hand, the energy of secondary radiation particles are necessarily lower than the incident runaway energy, thus reducing the overall energy of the runaway population. On the other hand, secondary radiation is emitted with broad energy and angular distributions, such that the majority of secondary electrons are not aligned with the magnetic field for parallel electric field acceleration. Future studies on runaway orbit simulations will quantify the role of these secondary electrons on plasma current decay.

    As a final note, we observe that generation of many relativistic or suprathermal electrons can help carry the current and thus slow down the current dissipation process. That is, producing a secondary population of current-carrying electrons may slow the overall current decay, reducing electromagnetic stresses associated with rapid current decay. Given that rapid termination of the runaway current ($\tau\sim 10~\mathrm{\mu s}$) could induce large mechanical stresses in the superconducting magnets of a tokamak, the effect of secondary radiation emission to mitigate this may prove useful.


    \section{Conclusions}
    \label{sec:conc}

    We have assessed the efficacy of a stand-off runaway termination scheme by tungsten particulate injection using MCNP simulations. For runaways incident with energies from 1 to 10 MeV, tungsten particulates effectively mitigate the runaways from energetics and kinematics points of view. Energetically, tungsten particulates can remove 99\% of energy from the runaway current beam on a timescale of 4--9 $\mu$s. Kinematically, these particulates will terminate 99\% of the initial runaway population on a timescale of 3--4 $\mu$s by pitch-angle scattering and absorption mechanisms. By either metric, 99\% of the runaway beam can be terminated before the onset of particulate melting, thus preventing any splattering of the first wall. Furthermore, the simulations show that runaway reseeding by secondary radiation from tungsten particulates is not a serious challenge for runaway termination. Secondary radiation is emitted with lower particle energies than the incident runaway and with a broad, nearly uniform angular distribution such that the majority of secondary electrons will not experience runaway acceleration.

    Future studies will include plasma physics simulations to study the orbital trajectories of runaways after strong pitch-angle scattering, particularly to verify if the runaway impact distribution across the first wall is sufficiently broadened to avoid serious damage as expected. More broadly, experimental studies of runaway termination by tungsten particulates in tokamak reactors are required to validate the simulation results. Overall, stand-off runaway termination by tungsten particulates is a promising scheme for last-ditch defense against runaway final impact and the resulting catastrophic damage.


    \ack This work is funded by the U.S. Department of Energy Office of Fusion Energy Sciences (DOE-FES) under the Tokamak Disruption Simulation (TDS) Scientific Discovery through Advanced Computing (SciDAC) project at Los Alamos National Laboratory (LANL) under Contract No. 89233218CNA000001.


    \section*{References}
    \bibliographystyle{unsrt}
    \bibliography{tungsten.bib}

\begin{thebibliography}{10}

\bibitem{ITER-2018}
{ITER} Organization.
\newblock {ITER} research plan within the staged approach (level {III} -
  provisional version).
\newblock Technical Report ITR-18-003, {ITER} Organization, St.
  Paul-lez-Durance, France, 2018.

\bibitem{RodriguezFernandez-2022-NuclFus}
P.~Rodriguez-Fernandez, A.J. Creely, M.J. Greenwald, D.~Brunner, S.B.
  Ballinger, C.P. Chrobak, D.T. Garnier, R.~Granetz, Z.S. Hartwig, N.T. Howard,
  J.W. Hughes, J.H. Irby, V.A. Izzo, A.Q. Kuang, Y.~Lin, E.S. Marmar, R.T.
  Mumgaard, C.~Rea, M.L. Reinke, V.~Riccardo, J.E. Rice, S.D. Scott, B.N.
  Sorbom, J.A. Stillerman, R.~Sweeney, R.A. Tinguely, D.G. Whyte, J.C. Wright,
  and D.V. Yuryev.
\newblock Overview of the {SPARC} physics basis towards the exploration of
  burning-plasma regimes in high-field, compact tokamaks.
\newblock {\em Nuclear Fusion}, 62(4):042003, 2022.

\bibitem{Martin-Solis-2017-NuclFus}
J.R. Martín-Solís, A.~Loarte, and M.~Lehnen.
\newblock Formation and termination of runaway beams in {ITER} disruptions.
\newblock {\em Nuclear Fusion}, 57(6):066025, 2017.

\bibitem{Breizman-2019-NuclFus}
Boris~N. Breizman, Pavel Aleynikov, Eric~M. Hollmann, and Michael Lehnan.
\newblock Physics of runaway electrons in tokamaks.
\newblock {\em Nuclear Fusion}, 59:083001, 2019.

\bibitem{McDevitt-2019-PPCF}
Christopher~J McDevitt, Zehua Guo, and Xian-Zhu Tang.
\newblock Avalanche mechanism for runaway electron amplification in a tokamak
  plasma.
\newblock {\em Plasma Physics and Controlled Fusion}, 61:054008, 2019.

\bibitem{Vallhagen-2020-JPlasPhys}
Oskar Vallhagen, Ola Embreus, Istvan Pusztai, Linnea Hesslow, and T{\"u}nde
  F{\"u}l{\"o}p.
\newblock Runaway dynamics in the {DT} phase of {ITER} operations in the
  presence of massive material injection.
\newblock {\em Journal of Plasma Physics}, 86(4):475860401, 2020.

\bibitem{Rosenbluth-1997-NuclFus}
M.N. Rosenbluth and S.V. Putvinski.
\newblock Theory for avalanche of runaway electrons in tokamaks.
\newblock {\em Nuclear Fusion}, 37(10):1355--1362, 1997.

\bibitem{Reux-2015-NuclFus}
C.~Reux, V.~Plyusnin, B.~Alper, D.~Alves, B.~Bazylev, E.~Belonohy, A.~Boboc,
  S.~Brezinsek, I.~Coffey, J.~Decker, P.~Drewelow, S.~Devaux, P.C. de~Vries,
  A.~Fil, S.~Gerasimov, L.~Giacomelli, S.~Jachmich, E.M. Khilkevitch,
  V.~Kiptily, R.~Koslowski, U.~Kruezi, M.~Lehnen, I.~Lupelli, P.J. Lomas,
  A.~Manzanares, A.~Martin~De Aguilera, G.F. Matthews, J.~Mlynář, E.~Nardon,
  E.~Nilsson, C.~Perez von Thun, V.~Riccardo, F.~Saint-Laurent, A.E. Shevelev,
  G.~Sips, C.~Sozzi, and {JET contributors}.
\newblock Runaway electron beam generation and mitigation during disruptions at
  {JET-ILW}.
\newblock {\em Nuclear Fusion}, 55(9):093013, 2015.

\bibitem{Matthews-2016-PhysScr}
G.F. Matthews, B.~Bazylev, A.~Baron-Wiechec, J.~Coenen, K.~Heinola, V.~Kiptily,
  H.~Maier, C.~Reux, V.~Riccardo, F.~Rimini, G.~Sergienko, V.~Thompson,
  A.~Widdowson, and {JET Contributors}.
\newblock Melt damage to the {JET} {ITER}-like {Wall} and divertor.
\newblock {\em Physica Scripta}, T167:014070, 2016.

\bibitem{Coburn-2022-NuclFus}
J.~Coburn, M.~Lehnen, R.A. Pitts, G.~Simic, F.J. Artola, E.~Thorén,
  S.~Ratynskaia, K.~Ibano, M.~Brank, L.~Kos, R.~Khayrutdinov, V.E. Lukash,
  B.~Stein-Lubrano, E.~Matveeva, and G.~Pautasso.
\newblock Energy deposition and melt deformation on the {ITER} first wall due
  to disruptions and vertical displacement events.
\newblock {\em Nuclear Fusion}, 62(1):016001, 2022.

\bibitem{Lehnen-2015-JNuclMater}
M.~Lehnen, K.~Aleynikova, P.B. Aleynikov, D.J. Campbell, P.~Drewelow, N.W.
  Eidietis, Yu. Gasparyan, R.S. Granetz, Y.~Gribov, N.~Hartmann, V.A. Izzo,
  S.~Jachmich, S.-H. Kim, M.~Kočan, H.R. Koslowski, D.~Kovalenko, U.~Kruezi,
  A.~Loarte, S.~Maruyama, G.F. Matthews, P.B. Parks, G.~Pautasso, R.A. Pitts,
  C.~Reux, V.~Riccardo, R.~Roccella, J.A. Snipes, A.J. Thornton, and P.C. {de
  Vries}.
\newblock Disruptions in {ITER} and strategies for their control and
  mitigation.
\newblock {\em Journal of Nuclear Materials}, 463(1):39--48, 2015.

\bibitem{Shiraki-2018-NuclFus}
D.~Shiraki, N.~Commaux, L.R. Baylor, C.M. Cooper, N.W. Eidietis, E.M. Hollmann,
  C.~Paz-Soldan, S.K. Combs, and S.J. Meitner.
\newblock Dissipation of post-disruption runaway electron plateaus by shattered
  pellet injection in {DIII-D}.
\newblock {\em Nuclear Fusion}, 58:056006, 2018.

\bibitem{Reux-2021-PhysRevLett}
C\'edric Reux, Carlos Paz-Soldan, Pavel Aleynikov, Vinodh Bandaru, Ondrej
  Ficker, Scott Silburn, Matthias Hoelzl, Stefan Jachmich, Nicholas Eidietis,
  Michael Lehnen, Sundaresan Sridhar, and JET contributors.
\newblock Demonstration of safe termination of megaampere relativistic electron
  beams in tokamaks.
\newblock {\em Physical Review Letters}, 126:175001, 2021.

\bibitem{PazSoldan-2021-NuclFus}
C.~Paz-Soldan, C.~Reux, K.~Aleynikova, P.~Aleynikov, V.~Bandaru, M.~Beidler,
  N.~Eidietis, Y.Q. Liu, C.~Liu, A.~Lvovskiy, S.~Silburn, L.~Bardoczi,
  L.~Baylor, I.~Bykov, D.~Carnevale, D.~Del-Castillo Negrete, X.~Du, O.~Ficker,
  S.~Gerasimov, M.~Hoelzl, E.~Hollmann, S.~Jachmich, S.~Jardin, E.~Joffrin,
  C.~Lasnier, M.~Lehnen, E.~Macusova, A.~Manzanares, G.~Papp, G.~Pautasso,
  Z.~Popovic, F.~Rimini, D.~Shiraki, C.~Sommariva, D.~Spong, S.~Sridhar,
  G.~Szepesi, C.~Zhao, the DIII-D~Team, and JET Contributors.
\newblock A novel path to runaway electron mitigation via deuterium injection
  and current-driven {MHD} instability.
\newblock {\em Nuclear Fusion}, 61(11):116058, 2021.

\bibitem{McDevitt-2023-PhysRevE}
Christopher~J. McDevitt and Xian-Zhu Tang.
\newblock Runaway electron current reconstitution after a nonaxisymmetric
  magnetohydrodynamic flush.
\newblock {\em Phys. Rev. E}, 108(4):L043201, 2023.

\bibitem{Boozer-2011-PPCF}
Allen~H Boozer.
\newblock Two beneficial non-axisymmetric perturbations to tokamaks.
\newblock {\em Plasma Physics and Controlled Fusion}, 53(8):084002, 2011.

\bibitem{Izzo-2022-NuclFus}
V.A. Izzo, I.~Pusztai, K.~Särkimäki, A.~Sundström, D.T. Garnier,
  D.~Weisberg, R.A. Tinguely, C.~Paz-Soldan, R.S. Granetz, and R.~Sweeney.
\newblock Runaway electron deconfinement in {SPARC} and {DIII-D} by a passive
  {3D} coil.
\newblock {\em Nuclear Fusion}, 62(9):096029, 2022.

\bibitem{Maviglia-2022-FusEngDes}
Francesco Maviglia, Christian Bachmann, Gianfranco Federici, Thomas Franke,
  Mattia Siccinio, Raffaele Albanese, Roberto Ambrosino, Wayne Arter, Roberto
  Bonifetto, Giuseppe Calabrò, Riccardo {De Luca}, Luigi E.~Di Grazia,
  Emiliano Fable, Pierluigi Fanelli, Alessandra Fanni, Mehdi Firdaouss,
  Jonathan Gerardin, Riccardo Lombroni, Massimiliano Mattei, Matteo Moscheni,
  William Morris, Gabriella Pautasso, Sergey Pestchanyi, Giuseppe Ramogida,
  Maria~Lorena Richiusa, Giuliana Sias, Fabio Subba, Fabio Villone, Jeong-Ha
  You, and Zsolt Vizvary.
\newblock Integrated design strategy for {EU-DEMO} first wall protection from
  plasma transients.
\newblock {\em Fusion Engineering and Design}, 117:113067, 2022.

\bibitem{Tang-2010-JFusErg}
X.~Z. Tang and G.~L. Delzanno.
\newblock Dust divertor for a tokamak reactor.
\newblock {\em Journal of Fusion Energy}, 29:407--411, 2010.

\bibitem{MCNP6-3}
Joel~Aaron Kulesza, Terry~R. Adams, Jerawan~Chudoung Armstrong, Simon~R.
  Bolding, Forrest~Brooks Brown, Jeffrey~S. Bull, Timothy~Patrick Burke,
  Alexander~Rich Clark, Robert~Arthur Forster, III, Jesse~Frank Giron,
  Tristan~Sumner Grieve, Colin~James Josey, Roger~Lee Martz, Gregg~Walter
  McKinney, Eric~J. Pearson, Michael~Evan Rising, Clell~Jeffrey Solomon, Jr.,
  Sriram Swaminarayan, Travis~John Trahan, Stephen~Christian Wilson, and
  Anthony~J. Zukaitis.
\newblock {MCNP\textsuperscript{\textregistered}} code version 6.3.0 theory \&
  user manual.
\newblock Technical Report LA-UR-22-30006, Los Alamos National Laboratory, Los
  Alamos, NM, USA,, 2022.

\bibitem{Berger-1963-MethCompPhys}
Martin~J. Berger.
\newblock {M}onte {C}arlo calculation of the penetration and diffusion of fast
  charged particles.
\newblock In B.~Alder, S.~Fernbach, and M.~Rotenberg, editors, {\em Methods in
  Computational Physics}, volume~1, pages 135--215. Academic Press, New York,
  NY, USA, 1963.

\bibitem{Hughes-2014-PNST}
H.~Grady Hughes.
\newblock Recent developments in low-energy electron/photon transport for
  {MCNP6}.
\newblock {\em Progress in Nuclear Science and Technology}, 4:454--458, 2014.

\bibitem{Matura-2022-IAEA}
Pascal Matura, Stefano Signetti, Stefan Moder, Luis Sandoval, Nathana{\"e}l
  Durr, Erkai Watson, Dilara Mert, and Markus B{\"u}ttner.
\newblock Modelling and simulation of the pellet shattering process related to
  the {SPI} technology for the {ITER} {DMS}.
\newblock Second Technical Meeting on Plasma Disruptions and their Mitigation,
  18-22 July 2022, Saint-Paul-lez-Durance, France, 2022.

\bibitem{Berger-1988-MCTrans}
Martin~J. Berger.
\newblock Electron stopping powers for transport calculations.
\newblock In T.M. Jenkins, W.R. Nelson, and A.~Rindi, editors, {\em {Monte
  Carlo} Transport of Electrons and Photons}, chapter~3, pages 57--80. Plenum
  Press, New York, NY, 10013, United States of America, 1988.

\bibitem{Seltzer-1988-MCTrans}
Stephen~M. Seltzer.
\newblock Cross sections for bremstrahhlung production and electron-impact
  ionization.
\newblock In Theodore~M. Jenkins, Walter~R. Nelson, and Alessandro Rindi,
  editors, {\em {Monte Carlo} Transport of Electrons and Photons}, chapter~4,
  pages 81--114. Plenum Press, New York, NY, 10013, United States of America,
  1988.

\bibitem{Hughes-2017-ICRS}
H.~Grady Hughes.
\newblock Improvements in electron-photon-relaxation data for {MCNP6}.
\newblock {\em EPJ Web of Conferences}, 153:06009, 2017.

\bibitem{Cullen-2014-EPICS}
Dermott~E. Cullen.
\newblock {EPICS2014}: Electron photon interaction cross sections (version
  2014).
\newblock Technical Report IAEA-NDS-218, International Atomic Energy Agency -
  Nuclear Data Section, Vienna, Austria, 2014.

\bibitem{Guo-2017-PPCF}
Zehua Guo, Christopher~J McDevitt, and Xian-Zhu Tang.
\newblock Phase-space dynamics of runaway electrons in magnetic fields.
\newblock {\em Plasma Physics and Controlled Fusion}, 59(4):044003, 2017.

\bibitem{MartinSolis-2015-PhysPlas}
J.R. Mart\'in-Sol\'is, A.~Loarte, and M.~Lehnen.
\newblock Runaway electron dynamics in tokamak plasmas with high impurity
  content.
\newblock {\em Physics of Plasmas}, 22(9):092512, 2015.

\end{thebibliography}


    \clearpage
    \onecolumn

    \renewcommand\figurename{Supplementary Figure}

    \article[\siname{}]{Supplementary material for the article entitled:}{Stand-off runaway electron beam termination by tungsten particulates for tokamak disruption mitigation}

    \author{Michael A. Lively,$^1$  \orcidlink{0000-0001-6511-9852}
        Danny Perez,$^1$        \orcidlink{0000-0003-3028-5249}
        Blas P. Uberuaga,$^2$   \orcidlink{0000-0001-6934-6219}
        Yanzeng Zhang$^1$       \orcidlink{0000-0002-1856-2701}
        and Xian-Zhu Tang$^1$   \orcidlink{0000-0002-4036-6643}}

    \address{$^1$ Theoretical Division, Los Alamos National Laboratory, Los Alamos, NM 87545, United States of America}
    \address{$^2$ Materials Science and Technology Division, Los Alamos National Laboratory, Los Alamos, NM 87545, United States of America}
    \ead{livelym@lanl.gov}

    \submitto{\NF}

    \-\newline


    \noindent Here, we provide energy and angular distributions of secondary radiation fluxes exiting a tungsten particulate after a runaway or gamma ray collision. We also provide these distributions for the primary gamma radiation flux for simulations in which the incident (source) particles were gamma rays. These data may be of interest to some readers, and also demonstrate the capabilities of our MCNP simulations to obtain this information, which are valuable as input to plasma physics simulations with a appropriate further processing.

    In each figure, the radial axis indicates the energy with which particles of interest exit the material, and the angular axis indicates the exit direction, where $0^\circ$ is the incident direction. Exiting fluxes are normalized against the incident runaway flux with values corresponding to the color bar at right. Incident runaway or gamma ray energies are indicated at the top-right of each plot.

    \clearpage

    \begin{figure*}[h!]
        \centering
        \begin{subfigure}{5.35in}
            \includegraphics[width=1.75in]{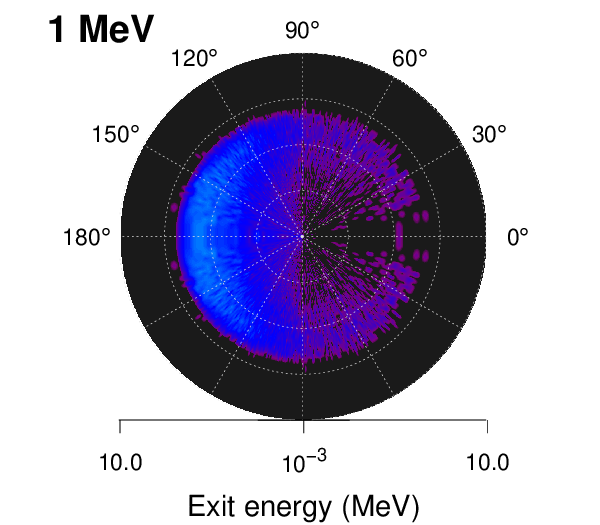}
            \includegraphics[width=1.75in]{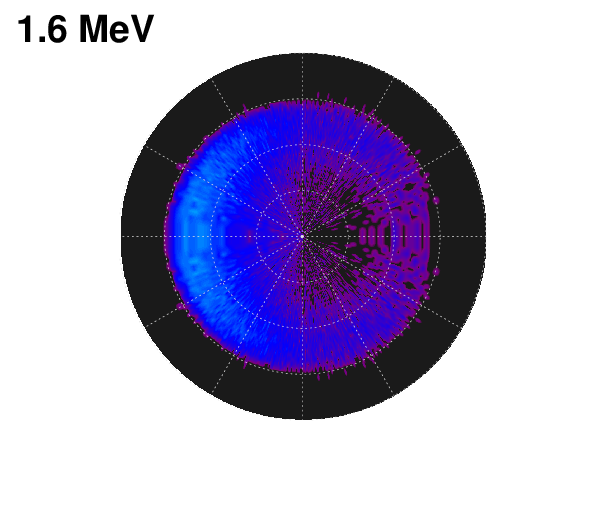}
            \includegraphics[width=1.75in]{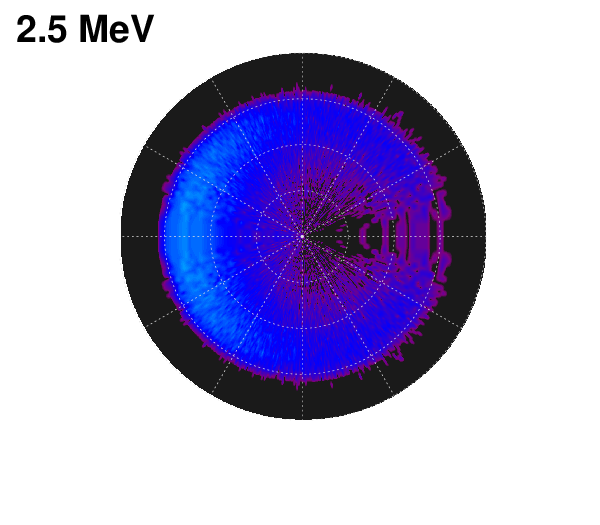}
            \\~\\
            \includegraphics[width=1.75in]{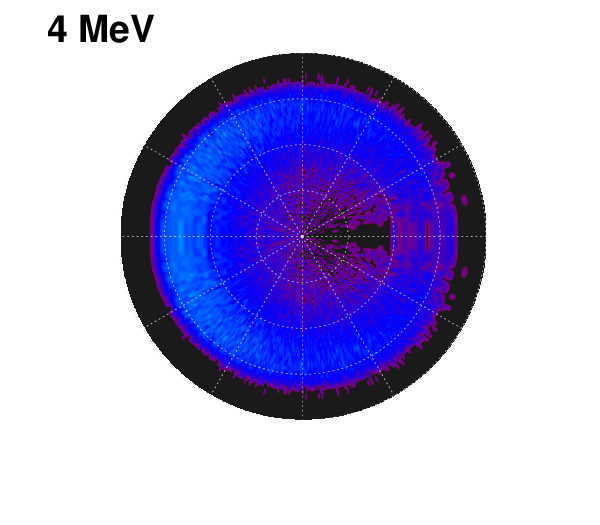}
            \includegraphics[width=1.75in]{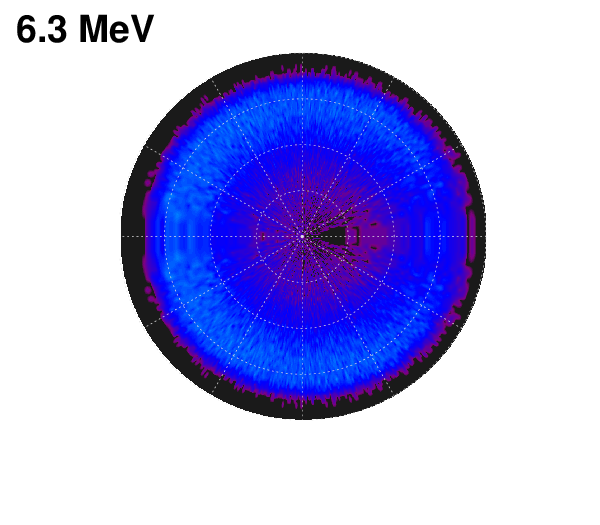}
            \includegraphics[width=1.75in]{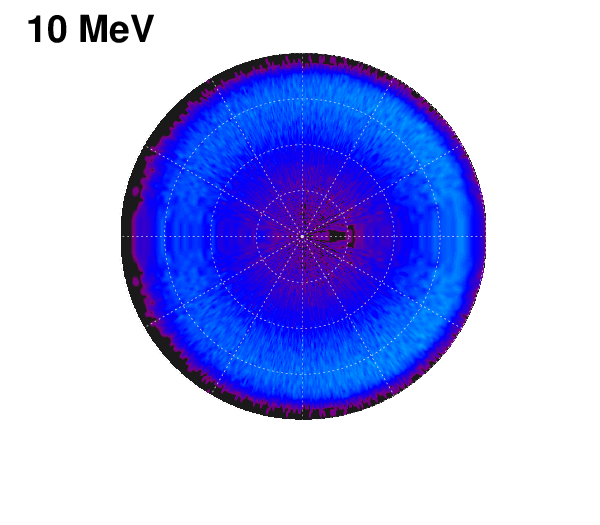}
        \end{subfigure}
        \begin{subfigure}{1in}
            \centering
            \includegraphics[height=3.25in]{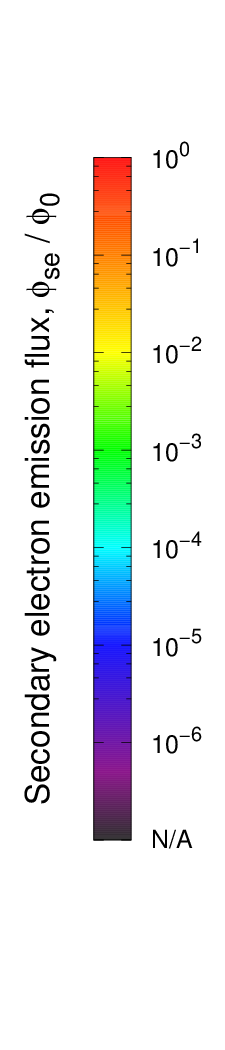}
        \end{subfigure}
        \caption{Energy and angle-resolved \underline{secondary electron emission} flux distributions resulting from a \underline{runaway} collision with a tungsten particulate.}
    \end{figure*}

    \-

    \begin{figure*}[h!]
        \centering
        \begin{subfigure}{5.35in}
            \includegraphics[width=1.75in]{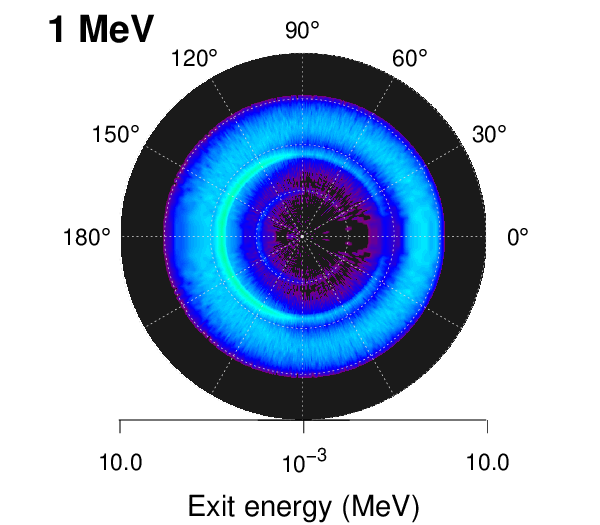}
            \includegraphics[width=1.75in]{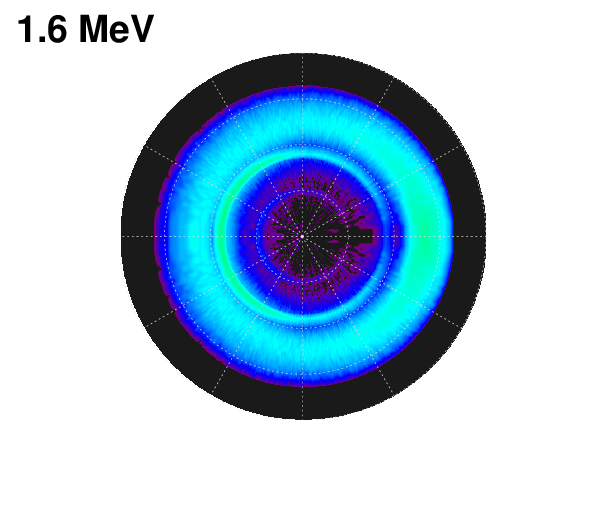}
            \includegraphics[width=1.75in]{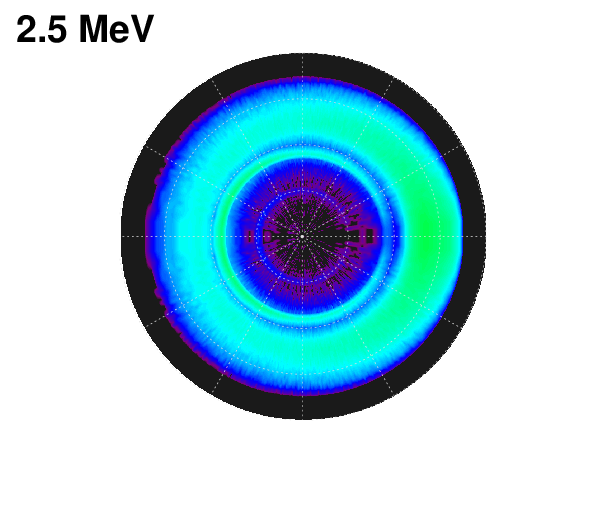}
            \\~\\
            \includegraphics[width=1.75in]{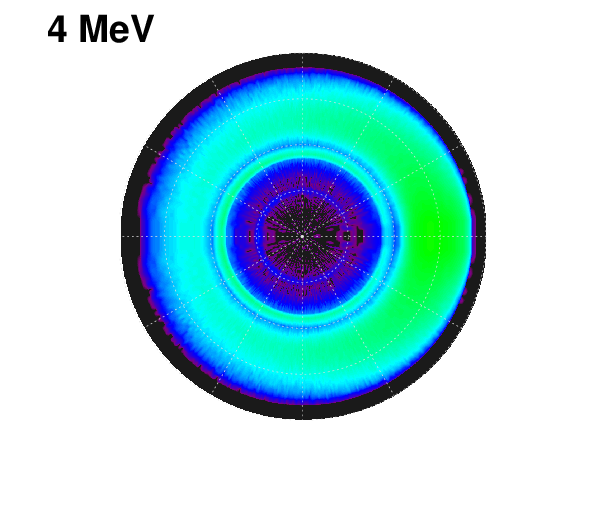}
            \includegraphics[width=1.75in]{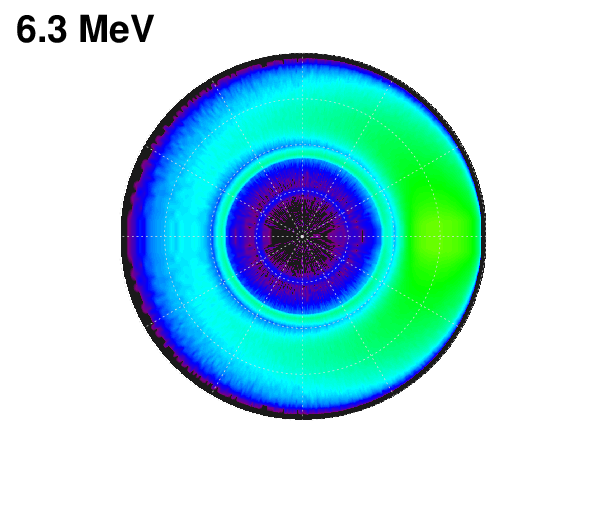}
            \includegraphics[width=1.75in]{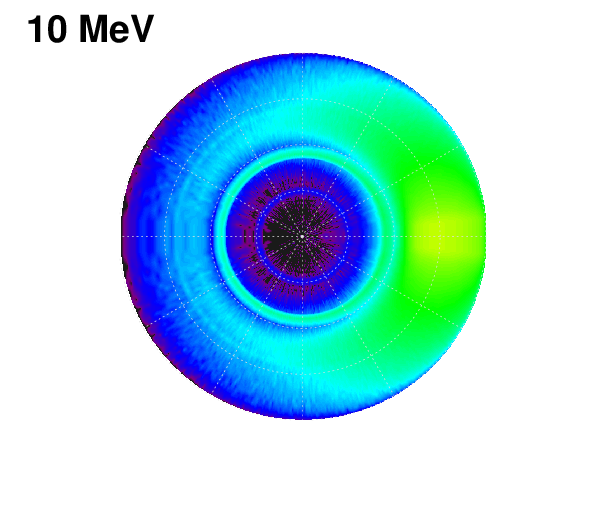}
        \end{subfigure}
        \begin{subfigure}{1in}
            \centering
            \includegraphics[height=3.25in]{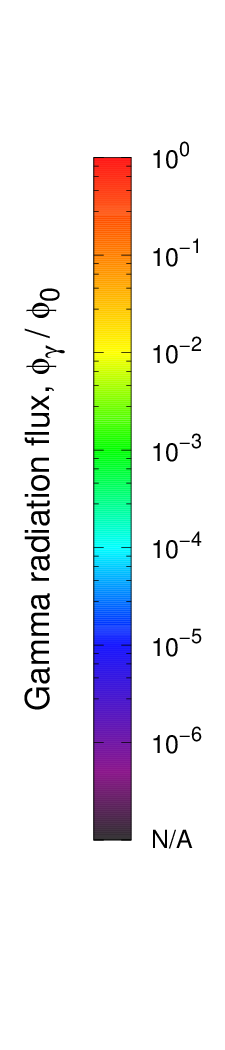}
        \end{subfigure}
        \caption{Energy and angle-resolved \underline{gamma radiation} flux distributions resulting from a \underline{runaway} collision with a tungsten particulate.}
    \end{figure*}

    \-

    \begin{figure*}[h!]
        \centering
        \begin{subfigure}{5.35in}
            \includegraphics[width=1.75in]{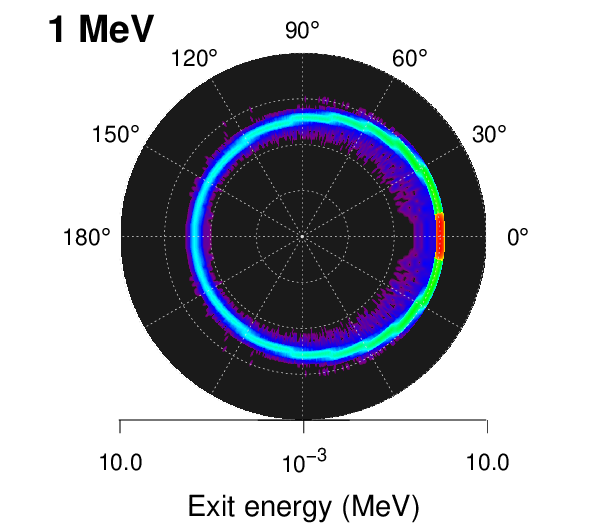}
            \includegraphics[width=1.75in]{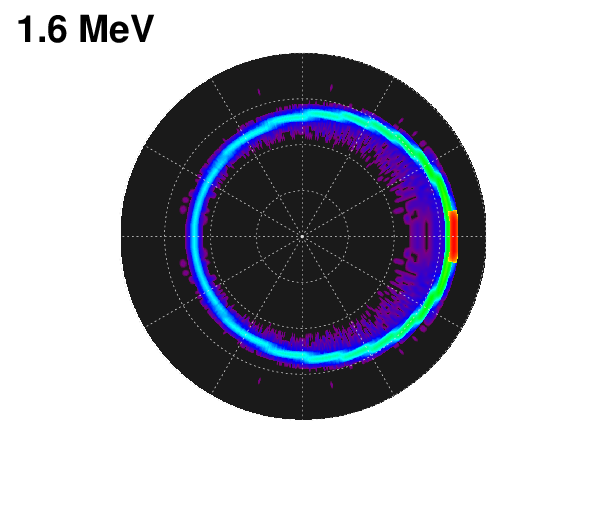}
            \includegraphics[width=1.75in]{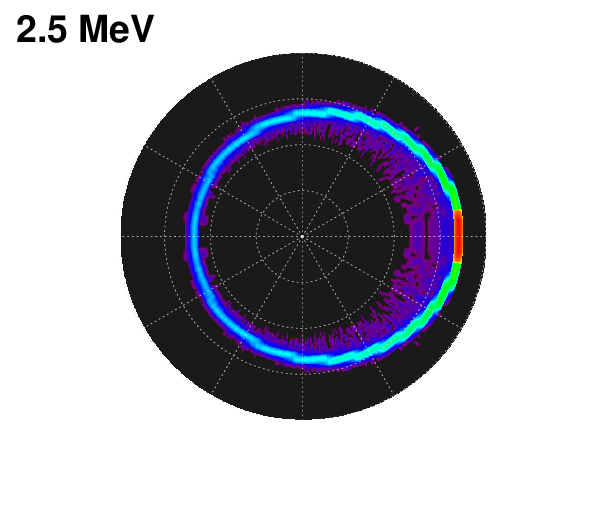}
            \\~\\
            \includegraphics[width=1.75in]{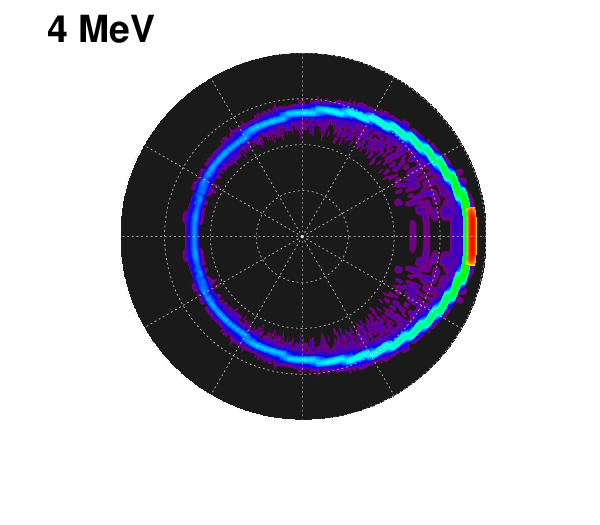}
            \includegraphics[width=1.75in]{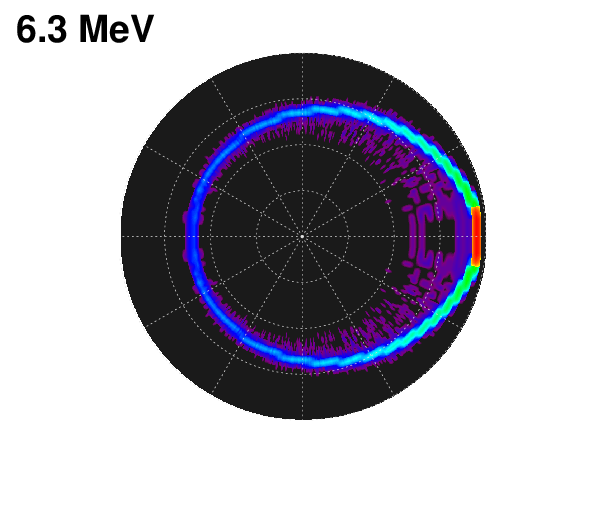}
            \includegraphics[width=1.75in]{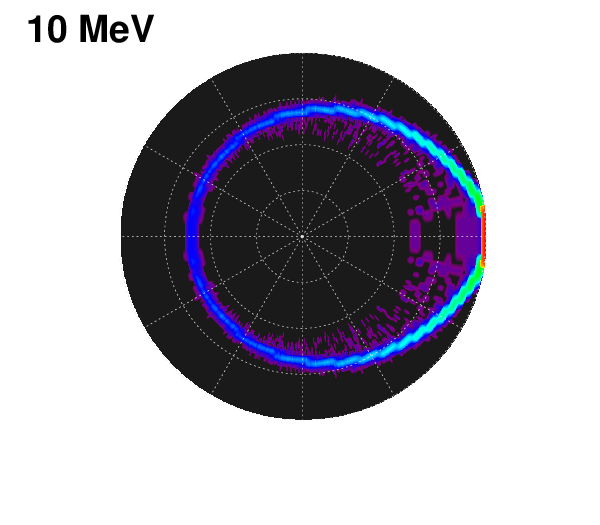}
        \end{subfigure}
        \begin{subfigure}{1in}
            \centering
            \includegraphics[height=3.25in]{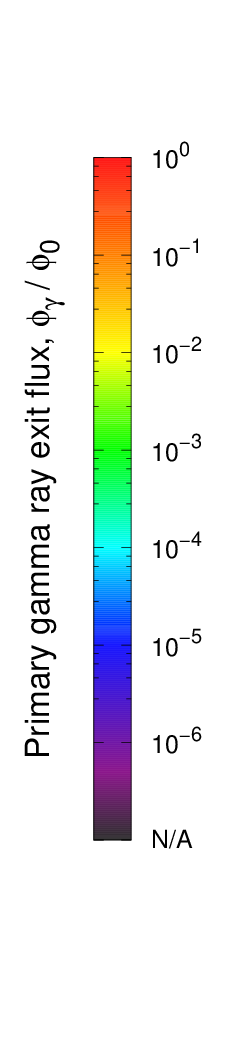}
        \end{subfigure}
        \caption{Energy and angle-resolved \underline{primary gamma ray} exit flux distributions resulting from a \underline{gamma ray} collision with a tungsten particulate.}
    \end{figure*}

    \-

    \begin{figure*}[h!]
        \centering
        \begin{subfigure}{5.35in}
            \includegraphics[width=1.75in]{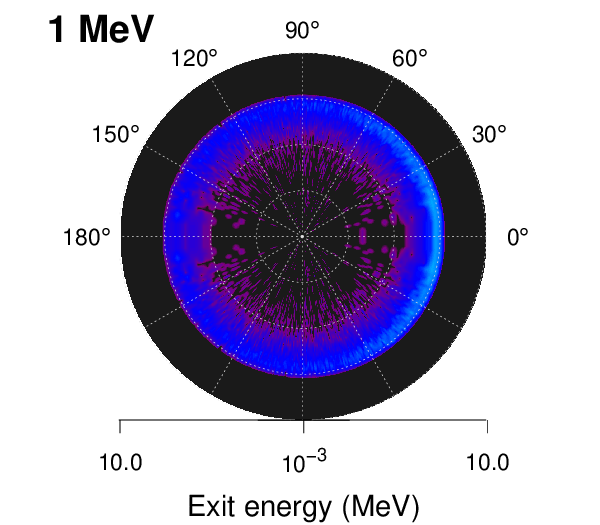}
            \includegraphics[width=1.75in]{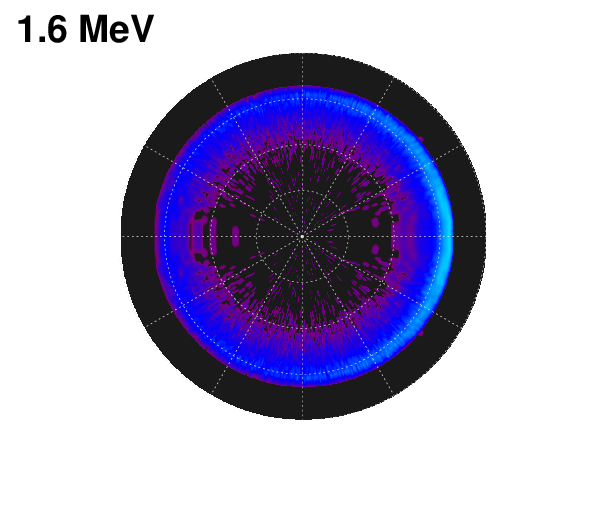}
            \includegraphics[width=1.75in]{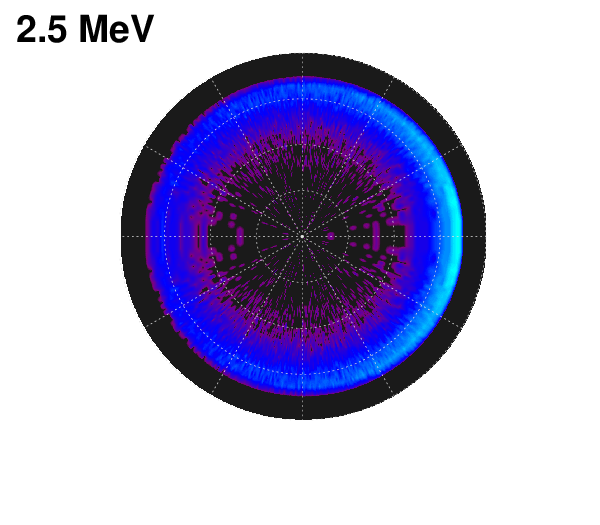}
            \\~\\
            \includegraphics[width=1.75in]{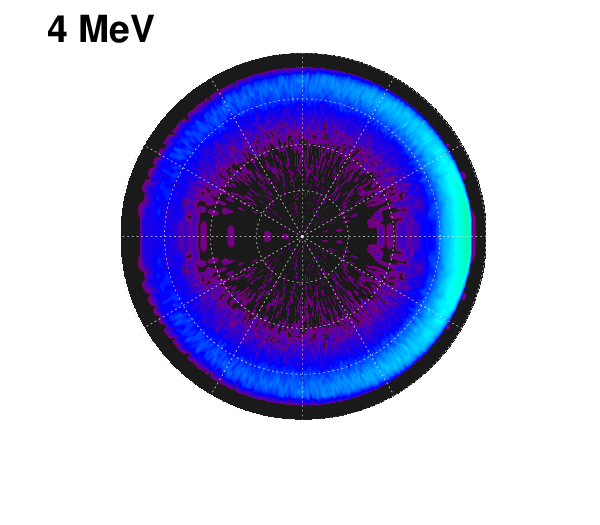}
            \includegraphics[width=1.75in]{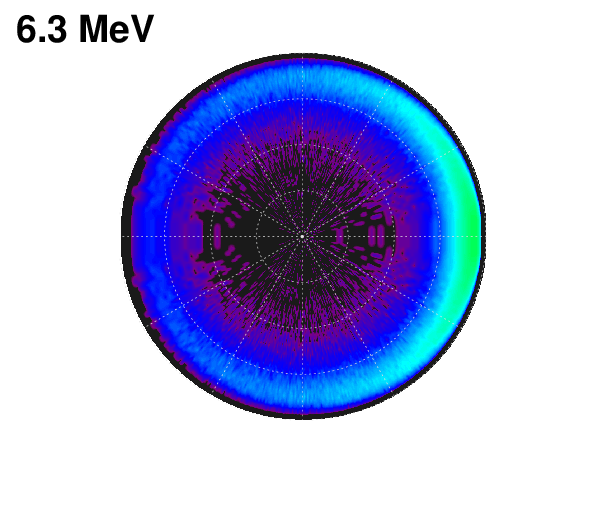}
            \includegraphics[width=1.75in]{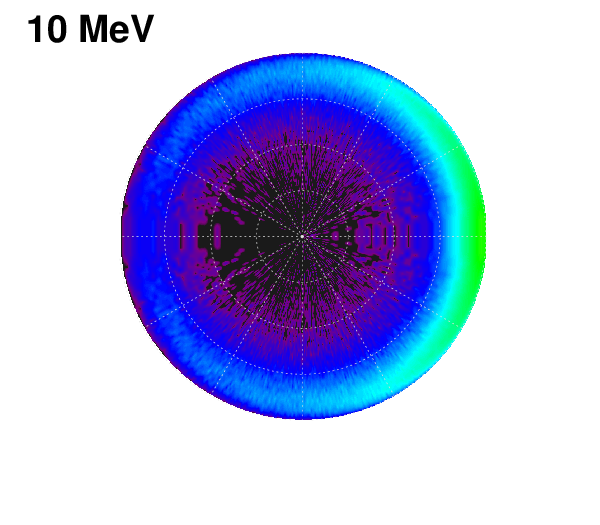}
        \end{subfigure}
        \begin{subfigure}{1in}
            \centering
            \includegraphics[height=3.25in]{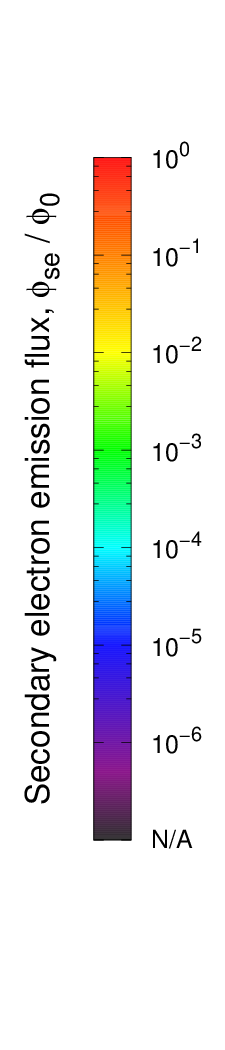}
        \end{subfigure}
        \caption{Energy and angle-resolved \underline{secondary electron emission} flux distributions resulting from a \underline{gamma ray} collision with a tungsten particulate.}
    \end{figure*}

    \-

    \begin{figure*}[h!]
        \centering
        \begin{subfigure}{5.35in}
            \includegraphics[width=1.75in]{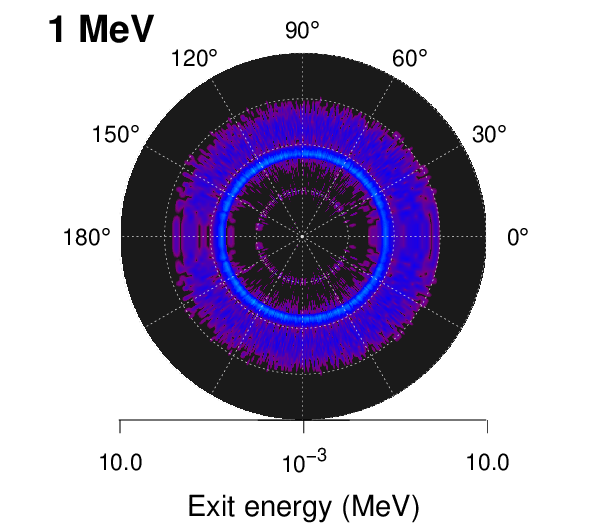}
            \includegraphics[width=1.75in]{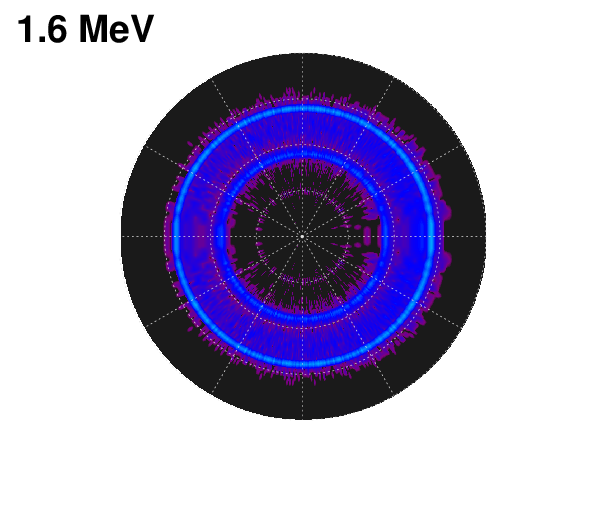}
            \includegraphics[width=1.75in]{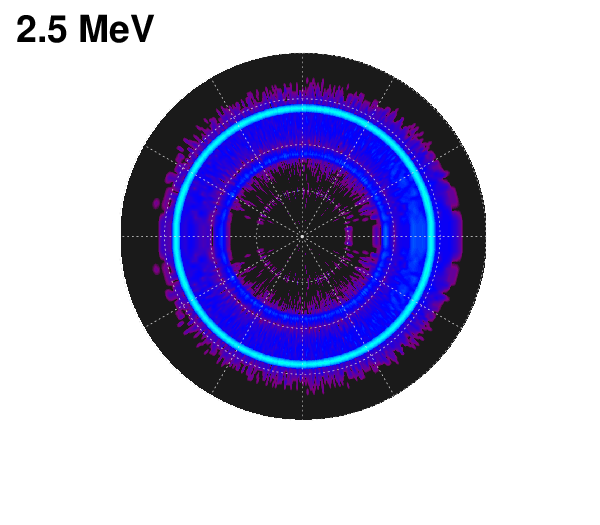}
            \\~\\
            \includegraphics[width=1.75in]{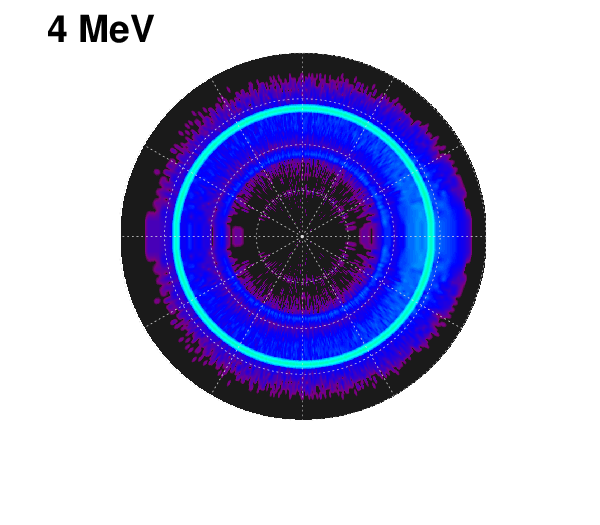}
            \includegraphics[width=1.75in]{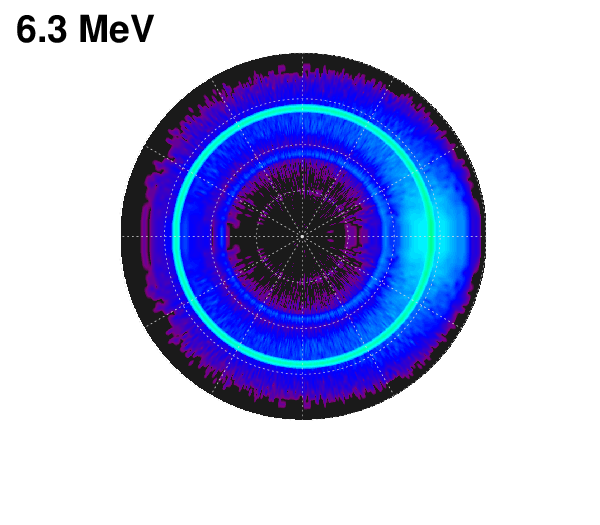}
            \includegraphics[width=1.75in]{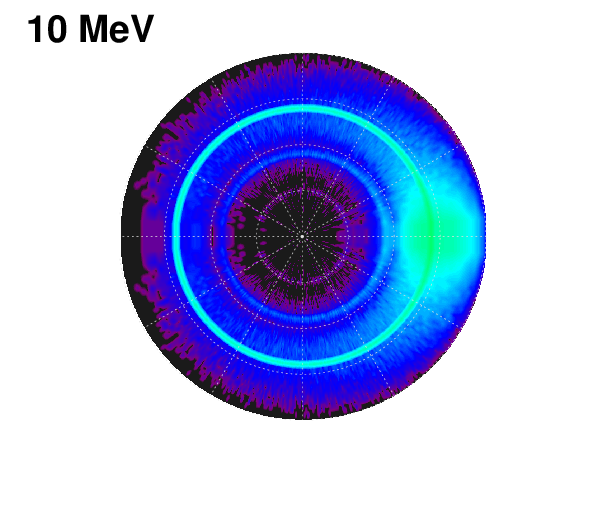}
        \end{subfigure}
        \begin{subfigure}{1in}
            \centering
            \includegraphics[height=3.25in]{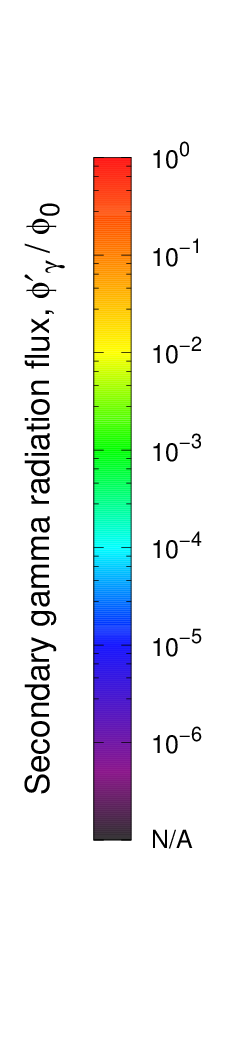}
        \end{subfigure}
        \caption{Energy and angle-resolved \underline{secondary gamma radiation} flux distributions resulting from a \underline{gamma ray} collision with a tungsten particulate.}
    \end{figure*}

\end{document}